\documentclass[useAMS,usenatbib]{mn2e}

\usepackage[english]{babel}
\usepackage[latin1]{inputenc}
\usepackage{amssymb}
\usepackage{graphicx}
\usepackage{color}
\usepackage{latexsym}
\usepackage{multirow}
\usepackage{txfonts}
\usepackage{amstext}
\def\LeP{\textit{Le PHARE}}
\def\mst{\mbox{$M_{\star}$}}
\def\Msun{\mbox{$M_\odot$}}
\def\lsim{\mathrel{\rlap{\lower3.5pt\hbox{\hskip0.5pt$\sim$}}     \raise0.5pt\hbox{$<$}}}                % less than or approx. symbol

\def\zs{\mbox{{$z_{\rm spec}$}}}
\def\zp{\mbox{{$z_{\rm phot}$}}}

\def\magapsix{\mbox{{\tt MAGAP\_6}}}
\title[Cooperative photo-z estimation]{A cooperative approach among methods for photometric redshifts estimation: an application to KiDS data}
\author[Cavuoti S. et al. ]{S.~Cavuoti$^{1}$\thanks{E-mail: stefano.cavuoti@gmail.com}, C.~Tortora$^{2}$, M.~Brescia$^{1}$, G.~Longo$^{3,4}$, M.~Radovich$^{5}$, N.~R.~Napolitano$^{1}$, \and V. Amaro$^{3}$, C. Vellucci$^{3}$, F.~La Barbera$^{1}$, F.~Getman$^{1}$, A.~Grado$^{1}$\\
$^{1}$Astronomical Observatory of Capodimonte - INAF, via Moiariello 16, I-80131 Napoli, Italy\\
$^{2}$Kapteyn Astronomical Institute, University of Groningen, P.O. Box 800, 9700 AV Groningen, the Netherlands\\
$^{3}$Department of Physics, University Federico II, Via Cinthia 6, I-80126 Napoli, Italy\\
$^{4}$California Institute of Technology-Center for data driven discovery, Pasadena CA-91125, USA\\
$^{5}$Astronomical Observatory of Padua, vicolo dell'Osservatorio 5, I-35122 Padova, Italy}
\begin{document}

\date{Accepted 07 December 2016. Received 9 November 2016; in original form 2 May 2016}

\pagerange{\pageref{firstpage}--\pageref{lastpage}} \pubyear{2016}

\maketitle

\label{firstpage}

\begin{abstract}
Photometric redshifts (photo-z's) are fundamental in galaxy surveys to address different topics, from gravitational lensing and dark matter distribution to galaxy evolution. The Kilo Degree Survey (KiDS), i.e. the ESO public survey on the VLT Survey Telescope (VST), provides the unprecedented opportunity to exploit a large galaxy dataset with an exceptional image quality and depth in the optical wavebands. Using a KiDS subset of about $25,000$ galaxies with measured spectroscopic redshifts, we have derived photo-z's using \textit{i}) three different empirical methods based on supervised machine learning, \textit{ii}) the Bayesian Photometric Redshift model (or BPZ), and \textit{iii}) a classical SED template fitting procedure (\LeP). We confirm that, in the regions of the photometric parameter space properly sampled by the spectroscopic templates, machine learning methods provide better redshift estimates, with a lower scatter and a smaller fraction of outliers. SED fitting techniques, however, provide useful information on the galaxy spectral type which can be effectively used to constrain systematic errors and to better characterize potential catastrophic outliers.
Such classification is then used to specialize the training of regression machine learning models, by demonstrating that a hybrid approach, involving SED fitting and machine learning in a single collaborative framework, can be effectively used to improve the accuracy of photo-z estimates.
\end{abstract}


\section{Introduction}

With the advent of modern multi-band digital sky surveys, photometric redshifts (photo-z's) have become crucial to provide redshift estimates for the large samples of galaxies which are required to tackle a variety of problems:
weak gravitational lensing to constrain dark matter and dark energy (\citealt{Kuijken+15_GL_KiDS}), the identification of galaxy clusters and groups (e.g. \citealt{Capozzi}; \citealt{Radovich+15_proc_clusters_KiDS}; \citealt{biviano2013}), the search of strong lensing (\citealt{Napolitano+15_proc_lensing_KiDS}) and ultra-compact galaxies (\citealt{Tortora+15_KiDS_compacts}), as well as the study of the mass function of galaxy clusters \citep{Annunziatella,Albrecht,Peacock,Keiichi}; to quote just a few.
Today, despite the initial skepticism \citep{Baum,Puschell,Koo,Loh}, two decades of continuous
improvements of photo-z techniques have led to such increase in accuracy that many ongoing
and planned surveys base their core science on photo-z measurements to fulfill their key scientific goals (e.g. \citealt{deJong+15_KIDS_paperI}; \citealt{masters2015}).

The evaluation of photo-z's is made possible by the existence of a complex correlation among the fluxes,
as measured by broad band photometry, the spectral types of the galaxies and their distance.
However, the search for the highly nonlinear function which maps the photometric parameter space into the redshift
one is far from trivial and can be performed in many different ways.
All existing methods can be divided into two main classes: theoretical and empirical.

\textit{Theoretical methods} use Spectral Energy Distribution (SED) templates derived either from observed galaxy spectra or from synthetic ones. Template based techniques are, on average, less accurate than empirical methods, but they are also free from the limitations imposed by the need of a training set. Moreover, SED fitting methods can be applied over a wide range of redshifts and intrinsic colors.
They rely, however, on using a set of galaxy templates that accurately maps the true distribution of galaxy spectral energy distributions (and their evolution with redshift), as well as on the assumption that the photometric calibration of the data is free from systematics. Finally, they also require a detailed understanding of how external factors, such as intergalactic and galactic extinctions, affect the final result.
The templates are then shifted to any redshift in a given range and convolved with the transmission curves of the filters to create a template set for the redshift estimators \citep{Koo2,aMassarotti,bMassarotti,Csabai,ilbert2006}.
Photometric redshifts can then be obtained by comparing observed galaxy fluxes in the $i^{th}$ photometric band with the library of reference fluxes, depending on (bounded by) redshift $z$ and on a set of parameters $T$, that account for galaxy spectral type. For each galaxy, a $\chi^2$ confidence test provides the values of $z$ and $T$ that minimize the flux residuals between observations and reference templates. A further improvement over the standard template methods was the introduction of magnitude priors defined within a Bayesian framework (BPZ; \citealt{Benitez}), which contributes to address important information on the galaxy types and expected shape of redshift distribution.

\textit{Empirical methods} use a Knowledge Base (KB hereafter) of objects with spectroscopically-measured redshifts as a training set to obtain an empirical correlation (i.e. the mapping function) between the photometric quantities and the redshift. Empirical methods have the advantage that they do not need accurate templates, because the training set is composed by real objects, which intrinsically include effects such as the filter bandpass and flux calibration, as well as reddening. However, these methods require that the KB must provide a good coverage of the photometric space, since unreliable redshift estimates are likely to  be obtained outside the color-redshift ranges covered by the KB \citep{masters2015,biviano2013,brescia2013,sanchez2014,brescia2015}.

Several estimators have been tested to determine the shape of the empirical mapping function, from linear or non-linear fitting (see e.g., \citealt{Connolly}),  to the use of machine learning algorithms such as Support Vector Machines \citep{chang2011}, Artificial Neural Networks \citep{pitts1943} and Instance-Based Learning \citep{russell2003}.
In recent times, several attempts to combine empirical and theoretical methods as well as other approaches, based on the combination or stacking of machine learning methods, have been discussed in literature, (\citealt{wolpert1992,carrasco2014,kim2015,beck2016,speagle2016,zitlau2016} and \citealt{fotopoulou2016}).

Blind tests of different methods to evaluate photo-z's have been performed in \cite{Hogg} on spectroscopic data from the Keck telescope on the Hubble Deep Field (HDF), in \cite{Hildebrandt2008} on spectroscopic data from the VIMOS VLT Deep Survey (VVDS) and the FORS Deep Field (FDF, \citealt{Noll}) on the sample of luminous red galaxies from the SDSS-DR6. A significant advance in comparing different methods was proposed in \cite{Hildebrandt2010}, through the so-called PHAT (PHoto-z Accuracy Testing) contest, which adopted the ``black-box'' approach which is typical of proper benchmarking. They performed a homogeneous comparison of the performances, focusing the analysis on the photo-z methods themselves, and setting an effective standard for the assessment of photo-z accuracy.

In \cite{Cavuoti+15_KIDS_I} we applied an empirical method based on machine learning, i.e. the Multi Layer Perceptron with Quasi Newton Algorithm (MLPQNA, \citealt{cavuoti2012}; \citealt{brescia2013, brescia2014, brescia2015}) to a dataset of galaxies extracted from the Kilo Degree Survey (KiDS). The KiDS survey, thanks to the large area covered ($1500$ sq. deg. at the end of the survey), the good seeing ($\sim 0.7''$ median FWHM in the r-band) and pixel scale ($\sim 0.2\, \rm ''/pixel$), together with its depth (r-band limiting magnitude of $\sim 25$; $5\sigma$ at $SNR=5$), provides large datasets of galaxies with high photometric quality in the four optical bands \textit{u, g, r} and \textit{i}, very important for accurate galaxy morphology up to $z=0.5-0.6$.

In this work we apply five different photo-z techniques to the same KiDS dataset:
a) three empirical methods, namely: the above mentioned MLPQNA, the Random Forest (RF; \citealt{breiman2001}), and an optimization network based on the Levenberg-Marquardt learning rule (LEMON; \citealt{nocedal2006});
b) the \LeP\ SED template fitting \citep{arnouts1999,ilbert2006};
c) the Bayesian Photometric Redshift model \citep{Benitez}.
The final goal being to analyze the possibility to use these models in a cooperative way, in order to optimize the accuracy of photo-z estimation.

The matching with overlapping spectroscopic surveys such as SDSS \citep{ahn2012} and GAMA (Galaxy And Mass Assembly) \citep{driver2011} provides a uniform and well controlled dataset to investigate: a) which method provides the most accurate photo-z estimates, b) whether the combination of different methods might provide useful insights into the accuracy of the final estimates.

The paper is structured as follows. In Sect.~\ref{SEC:thedata} we present the data set. The methods used to evaluate photo-z's are summarized in Sect.~\ref{SEC:themethod}. In Sect.~\ref{SEC:exp} we describe the experiments and finally we discuss the results in Sect.~\ref{SEC:discussion}. Final remarks are outlined in Sect.~\ref{SEC:conclusions}.

\section{The data}\label{SEC:thedata}

KiDS is an optical survey \citep{deJong+15_KIDS_paperI}, carried out with the VST-OmegaCAM camera \citep{Kuijken11}, dedicated mainly to studies for gravitational lensing, galaxy evolution, searches for high-z quasars and galaxy clusters.
The KiDS data releases consist of tiles which are observed in \textit{u, g, r}, and \textit{i} bands. Data are processed using a distributed Oracle-based environment through the Astro-WISE (AW) optical pipeline \citep{McFarland+13}. Source extraction is performed using the algorithm KiDSCAT within the AW environment, where tile stacking, photometric calibration and astrometry are performed (see \citealt{deJong+15_KIDS_paperI}).

The sample of galaxies on which we performed our analysis is mostly extracted from the second data release of
KiDS (KiDS-DR2; \citealt{deJong+15_KIDS_paperI}) which contains $148$ tiles observed in all filters during the first two years of survey regular operations. We added $29$ extra tiles, not included in the DR2 at the time this was released, that will be part of the forthcoming KiDS data release, thus covering an area of $177$ square degrees.

We used the multi-band source catalogues, based on source detection in the r-band images. While magnitudes are measured in all filters, the star-galaxy separation, as well as the positional and shape parameters are derived from the r-band data only, which typically offers the best image quality and r-band seeing $\sim 0.65''$, thus providing the most reliable source positions and shapes. Critical areas such as saturated pixels, spikes, reflection halos, satellite tracks etc., are masked out, and galaxies are suitably flagged. Star/Galaxy separation is based on the {\tt CLASS\_STAR} (star classification) and SNR (signal-to-noise ratio) parameters provided by S-Extractor \citep{bertin1996}, see also \cite{labarbera2008} for further details about this procedure. We have retained sources with r-band S-Extractor $\tt FLAGS\_r < 4$, thus including objects that are very close together, very bright, with bad pixels, or blended. Further details about data reduction steps and catalogue extraction are provided in \cite{deJong+15_KIDS_paperI} and \cite{Tortora+15_KiDS_compacts}.

From the original catalogue of $\sim 22$ million sources, the Star/Galaxy separation leaves $\sim 12.2$ million of galaxies. After removing those galaxies which happen to fall in the masked regions, the final sample consisted of $\sim 7.6$ million galaxies.

Aperture photometry in the four \textit{ugri} bands, measured within several radii, was derived using S-Extractor. In this work we use magnitudes ${\tt MAGAP\_4}$ and ${\tt MAGAP\_6}$, measured within the apertures of diameters $4''$ and $6''$, respectively. These apertures were selected to reduce the effects of seeing and to minimize the contamination from mismatched sources. To correct for residual offsets in the photometric zero points, we used the SDSS as reference: for each KiDS tile and band we matched bright stars with the SDSS catalogue and computed the median difference between KiDS and SDSS magnitudes (\textit{psfMag}). For more details about data preparation and pre-processing see \cite{deJong+15_KIDS_paperI} and \cite{Cavuoti+15_KIDS_I}.

\subsection{Spectroscopic base}

In order to build the spectroscopic KB we cross-matched the KiDS data with the spectroscopic samples available in the GAMA data release $2$ (\citealt{driver2011}; \citealt{liske2015}) and SDSS-III data release $9$ (\citealt{ahn2012}; \citealt{bolton2012}; \citealt{chen2012}).

GAMA observes galaxies out to $z = 0.5$ and $r < 19.8$ (r-band petrosian magnitude), by reaching a spectroscopic completeness of $98\%$ for the main survey targets. It also provides information about the quality of the redshift determination by using the probabilistically defined normalized redshift quality scale $nQ$. Redshifts with $nQ  > 2$ were considered the most reliable (\citealt{driver2011}). For what concerns SDSS-III data, we used the low-z (LOWZ) and constant mass (CMASS) samples of the Baryon Oscillation Sky Survey (BOSS). The BOSS project obtains spectra (redshifts) for $1.5$ millions of luminous galaxies up to $z \sim 0.7$. The LOWZ sample consists of galaxies with $0.15 < z < 0.4$ with colors similar to luminous red galaxies (LRGs), selected by applying suitable cuts on magnitudes and colors to extend the SDSS LRG sample towards fainter magnitudes/higher redshifts (see e.g. \citealt{ahn2012}; \citealt{bolton2012}). The CMASS sample contains three times more galaxies than the LOWZ sample, and is designed to select galaxies with $0.4 < z < 0.8$. The rest-frame color distribution of the CMASS sample is significantly broader than that of the LRG one, thus CMASS contains a nearly complete sample of massive galaxies down to $\log \mst /\Msun \sim 11.2$. The faintest galaxies are at $r = 19.5$ for LOWZ and $i = 19.9$ for CMASS. Our matched spectroscopic sample is dominated by galaxies from GAMA ($46,598$ vs. $1,618$ from SDSS) at low-z ($z \lsim 0.4$), while SDSS galaxies dominate the higher redshift regime (out to $z \sim 0.7$), with $r < 22$.

\subsection{Knowledge base definition}

As a general rule, in order to avoid any possible misuse of the data, in each experiment we identified sources in the KB by adding a flag, specifying whether an object belongs to the training or test sets, respectively.

The detailed procedure adopted to obtain the two data sets used for the experiments is as follows:
\begin{itemize}
\item we excluded objects having low photometric quality (i.e. with flux error higher than one magnitude);
\item we removed all objects having at least one missing band (or labeled as Not-a-Number or NaN), thus obtaining the clean catalogue used to create the training and test sets, in which all required photometric and spectroscopic information is complete for all objects;
\item we performed a randomly shuffled splitting into a training and a blind test set, by using the $60\% / 40\%$ percentages, respectively;
\item we applied the following cuts on limiting magnitudes (see \citealt{cavuoti2015} for details):
\begin{itemize}
 \item MAGAP\_4\_u $\leq25.1 $
 \item MAGAP\_6\_u $\leq24.7 $
 \item MAGAP\_4\_g $\leq24.5 $
 \item MAGAP\_6\_g $\leq24.0 $
 \item MAGAP\_4\_r $\leq22.2 $
 \item MAGAP\_6\_r $\leq22.0 $
 \item MAGAP\_4\_i $\leq21.5 $
 \item MAGAP\_6\_i $\leq21.0 $
\end{itemize}
\item we selected objects with {\tt IMA\_FLAGS} equal to zero in the \textit{g, r} and \textit{i} bands (i.e. sources that have been flagged because located in proximity of saturated pixels, star haloes, image border or reflections, or within noisy areas, see \citealt{deJong+15_KIDS_paperI}). The \textit{u} band is not considered in such selection since the masked regions relative to this waveband are less extended than in the other three KiDS bands.
\end{itemize}

By applying all the specified steps, the final KB consisted of $15,180$ training and $10,067$ test objects. The cuts, of course, reduce the size of the final dataset for which reliable redshifts estimates can be obtained, see \cite{Cavuoti+15_KIDS_I} for more details.

We note that, as it is well known, empirical methods can be successfully applied only within the boundaries of the input parameter space, which is properly sampled by the knowledge base (cf. \citealt{masters2015}). In other words, any bias in the KB (e.g. photometric cuts, poorly represented groups of rare and peculiar objects, etc.) is reflected also in the results. This implies that the same prescriptions applied to the KB need to be applied also to the catalogues of objects for which we derive the photo-z's.

\section{The methods}\label{SEC:themethod}

In this section we shortly outline the empirical (MLPQNA, RF and LEMON), and the theoretical (\LeP, BPZ) methods which have been used for the
comparison which is discussed in the rest of the work.

\subsection{The Machine Learning models}

Among the methods which are made publicly available through the DAMEWARE (DAta Mining \& Exploration Web Application REsource; \citealt{brescia2014}) web-based infrastructure, we picked three machine learning models: the Random Forest (RF; \citealt{breiman2001}), and two versions of the Multi Layer Perceptron (MLP; \citealt{rosenblatt1961}), varying in terms of backward learning methods, i.e. the Quasi Newton Algorithm (QNA; \citealt{byrd1994}) and the Levenberg-Marquardt rule \citep{nocedal2006}, respectively.

Random Forest \citep{breiman2001} is an ensemble learning method for classification and regression. It is a collection of simple predictors, called decision trees, where each tree is capable of producing a response to a given pattern, by subdividing the data into smaller and smaller sets based on simple decisions. The main principle behind ensemble  methods is that a collection of ``weak learners'' can be joined to form a ``strong learner''. A Random Forest can then be considered as a meta estimator that fits a large number of decision trees on several sub-samples of the original training set and produces an average output. Such mechanism improves the predictive accuracy, with respect to the single decision tree, and keeps over-fitting under control.

LEMON (LEvenberg-Marquardt Optimization Network) is based on the modified Levenberg-Marquardt method, which makes use of the exact Hessian of the error function (and not of its linearized approximation). For networks with up to several hundreds of internal weights this algorithm is comparable with the QNA (often faster). But its main advantage is that it does not require any stopping criterion. This method almost always converges exactly to one of the minima of a function.

The MLPQNA model, i.e. a MLP implementation with learning rule based on the Quasi Newton Algorithm, belongs to the Newton's methods specialized to find the stationary point of a function through a statistical approximation of the Hessian of the training error, obtained by a cyclic gradient calculation. MLPQNA makes use of the known L-BFGS algorithm (Limited memory - Broyden Fletcher Goldfarb Shanno; \citealt{byrd1994}), originally designed for problems with a wide parameter space. The analytical details of the MLPQNA method, as well as its performances, have been extensively discussed elsewhere \citep{cavuoti2012,brescia2013,cavuoti2014b,cavuoti2015}.

Traditional supervised learning requires the KB to be split into training and test sets. The former is used to ``train'' the method, i.e. to infer the hidden relationship between the photometric information and the redshifts. The latter, instead, is used to evaluate - using a set of statistical estimators (see Sect.~\ref{SEC:stat}) - the goodness of the inferred law. To avoid biases, test and training sets are always required to have null intersection.

\subsection{\LeP\ SED fitting}

We use the standard SED fitting method, adopting the software \LeP\ \citep{arnouts1999,ilbert2006}. KiDS observed magnitudes are matched with those predicted from a set of spectral energy distributions (SEDs). Each SED template is redshifted in steps of $\delta z = 0.01$ and convolved with the four filter transmission curves. The following merit function (eq.~\ref{formula1}) is then minimized:
\begin{equation}
\chi^{2}(z,T,A) = \sum_{i=1}^{N_{f}} { \left( \frac{F_{\rm obs}^{f}-A\times F_{\rm pred}^{f}(z,T)}{\sigma_{obs}^{f}} \right)^2}
\label{formula1}
\end{equation}

\begin{figure}
\includegraphics[width=8 cm]{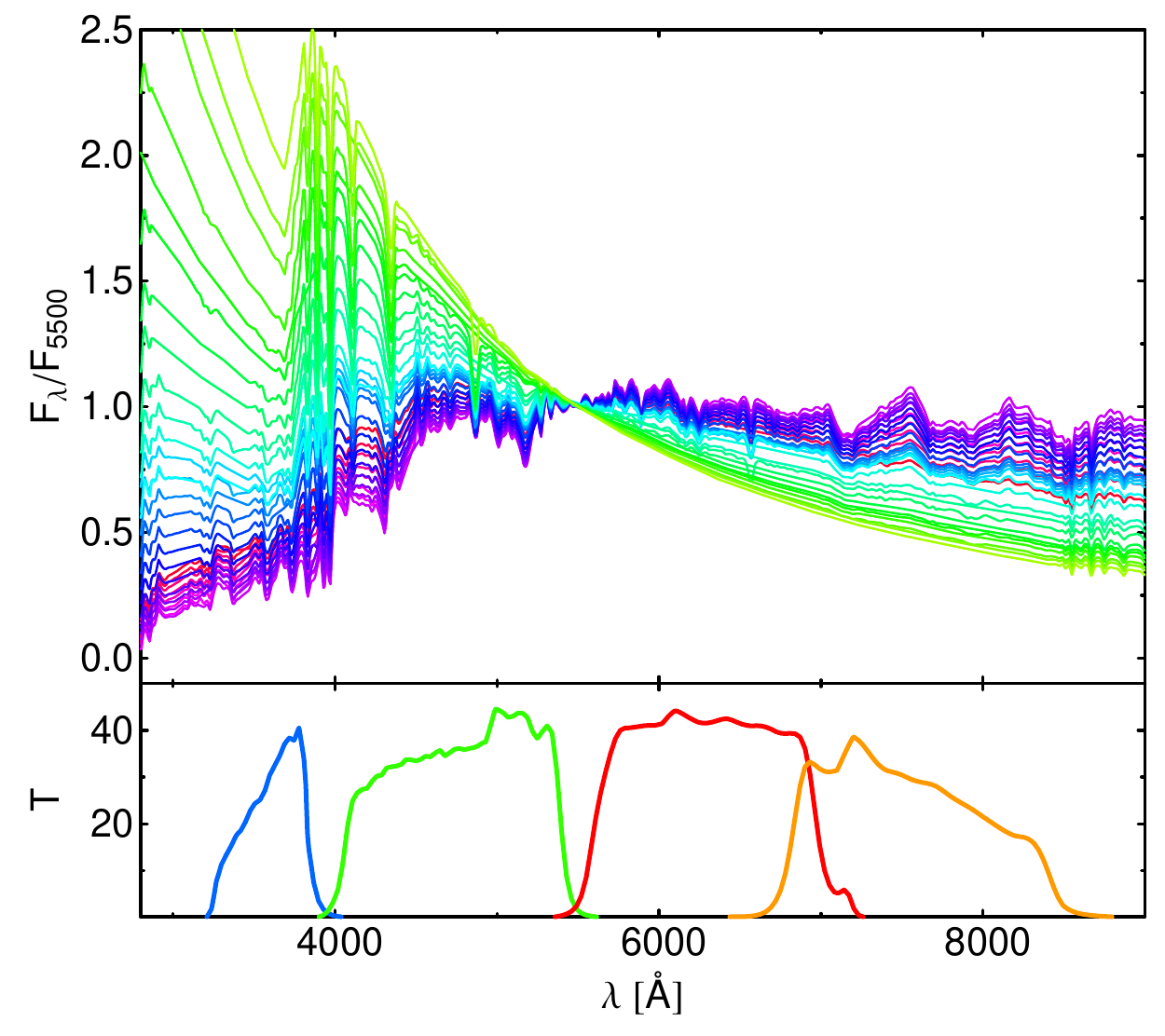}
\caption{SED templates. Flux normalized to the flux at $5000$ \AA\ vs. wavelength. Templates are taken from \citet{ilbert2006}, see text for details. Redder colors are for ellipticals, blue and green for spirals and irregulars, and finally the darker green is for starburst templates. In the bottom panel the KiDS filters are shown.}
\label{fig:CFHTLS_SED}
\end{figure}

\noindent where $F_{\rm pred}^{f}(z,T)$ is the flux predicted for a template T at redshift z. $F_{\rm obs}^{f}$ is the observed flux and $\sigma_{obs}^{f}$ the associated error derived from the observed magnitudes and errors. The index $f$ refers to the considered filter and $N_{f}=4$ is the number of filters. The photometric redshift is determined from the minimization of $\chi^{2}(z,T,A)$ varying the three free parameters z, T and the normalization factor A. As final products of the fitting procedure, the \LeP\ code provides two main results: a) the photometric redshift ($z=\zp$) and b) a galaxy spectral-type classification, based on the best-fitted template model T.

For the SED fitting experiments we used the \magapsix\ magnitudes in the \textit{u, g, r} and \textit{i} bands (and related $1\, \sigma$ uncertainties), corrected for galactic extinction using the map in \cite{Schlafly_Finkbeiner11}. As reference template set we adopted the $31$ SED models used for COSMOS photo-z's \citep{Ilbert+09} (see Fig.~\ref{fig:CFHTLS_SED}). The basic COSMOS library is composed by nine galaxy templates from \cite{Polletta+07}, which includes three SEDs of elliptical galaxies (\textit{E}) and five templates of spiral galaxies (\textit{S0, Sa, Sb, Sc, Sd}). These models are generated using the code GRASIL (\citealt{Silva+98}), providing a better joining of UV and mid-IR than those by \cite{coleman1980} used in \cite{ilbert2006}. Moreover, to reproduce very blue colors not accounted by the \cite{Polletta+07} models, $12$ additional templates using \cite{bruzual2003} models with starburst (\textit{SB}) ages ranging  from $3$ to $0.03$ Gyr are added. In order to improve the sampling of the redshift-color space and therefore the accuracy of the redshift measurements, the final set of $31$ spectra is obtained by linearly interpolating the original templates. We refer to it as the COSMOS library.
Internal galactic extinction can be also included as free parameter in the fitting procedure, using two different galactic extinction laws (\citealt{Prevot+84}; \citealt{Calzetti+00}), with $E_{BV} \leq 0.5$.

However, we followed the setup discussed in \cite{Ilbert+09}, i.e. we did not apply any galactic extinction correction for models redder than the \textit{Sc} templates; the galactic extinction curve provided by \cite{Prevot+84} is used for templates redder than \textit{SB3} model, while \cite{Calzetti+00} is adopted for those bluer (including \textit{SB3} template). Emission lines added to the templates were also implemented as discussed in \cite{Ilbert+09}. Finally, \LeP\ also provides an adaptive procedure, which calculates the shifts in the photometric zero-points. The fit is first performed on the training set: the redshift is fixed to its spectroscopic value and for each waveband the code calculates average shifts which minimize the differences between observed and predicted magnitudes. This procedure is applied iteratively until convergence is reached. The offsets are then applied to the observed magnitudes of galaxies in the test sample, and the minimization of the $\chi^2$ is performed.
We tried some preliminary experiments without imposing any constraint on the fitted models, finding that about $12 \%$ of the test sample would have estimated photometric redshifts larger than $1$, with most of them being catastrophic outliers. For this reason, by looking at the results for the test sample, we imposed the flat prior, derived from the training data only, on absolute magnitudes. In particular, we have forced the galaxies to have absolute \textit{i}-band magnitudes in the range $(-10,-26)$.

We also tested three different configurations: \textit{(i)} the fit of SED templates with no internal galactic extinction and no emission lines; \textit{(ii)} the fit of SED templates with no internal galactic extinction and no emission lines, but allowing the photometry zero-points to vary using the adaptive procedure in \LeP; \textit{(iii)} the fit of SED templates, using internal galactic extinction as a free parameter, adding emission lines, and using the adaptive procedure. The best results in terms of photo-z statistical performance (see Sect.~\ref{SEC:stat}) were obtained with the second configuration, which is referred hereafter as the SED fitting photo-z estimation result. The use of the spectral templates is an important part of this paper, and will be used in subsequent sections.

\subsection{Bayesian Photometric Redshifts}

BPZ (\citealt{Benitez}) is a Bayesian photo-z estimation based on a template fitting method. The BPZ library is composed \citep{Benitez2004} of four modified Coleman, Wu and Weedman types \citep{coleman1980}, plus two \cite{Kinney1996} starburst templates. The templates include emission lines, but no internal dust extinction. As recommended in the BPZ documentation,  we allowed BPZ to interpolate adjacent template pairs in the color space. If spectroscopic redshifts are available, BPZ computes the ratio of observed to model best-fit fluxes, thus allowing to derive a correction to the initial zero points.

The Bayesian approach adopted in BPZ combines the likelihood that a template fits the SED of a galaxy at a given redshift, with a prior defining the probability to find a galaxy of that type, as a function of magnitude and redshift. This allows to remove those solutions that would be selected if based only on the maximum likelihood, but are in disagreement with the observed distributions. In addition to the redshift and template, BPZ also provides for each galaxy the full redshift probability distribution, and a parameter (ODDS) which provides the reliability of the solution.

\subsection{Statistical estimators}\label{SEC:stat}
The results were evaluated using the following set of statistical estimators for the quantity $\Delta z = (\zs-\zp)/(1+\zs)$ on the objects in the blind test set:

\begin{itemize}
\item bias: defined as the mean value of the residuals $\Delta z$;
\item $\sigma$: the standard deviation of the residuals;
\item $\sigma_{68}$: the radius of the region that includes $68\%$ of the residuals close to 0;
\item NMAD: Normalized Median Absolute Deviation of the residuals, defined as $NMAD(\Delta z) = 1.48 \times Median (|\Delta z|)$;
\item fraction of outliers with $|\Delta z| > 0.15$.
\end{itemize}

\section{Combination of methods}\label{SEC:exp}

\begin{figure}
\centering
\includegraphics[width=0.45\textwidth]{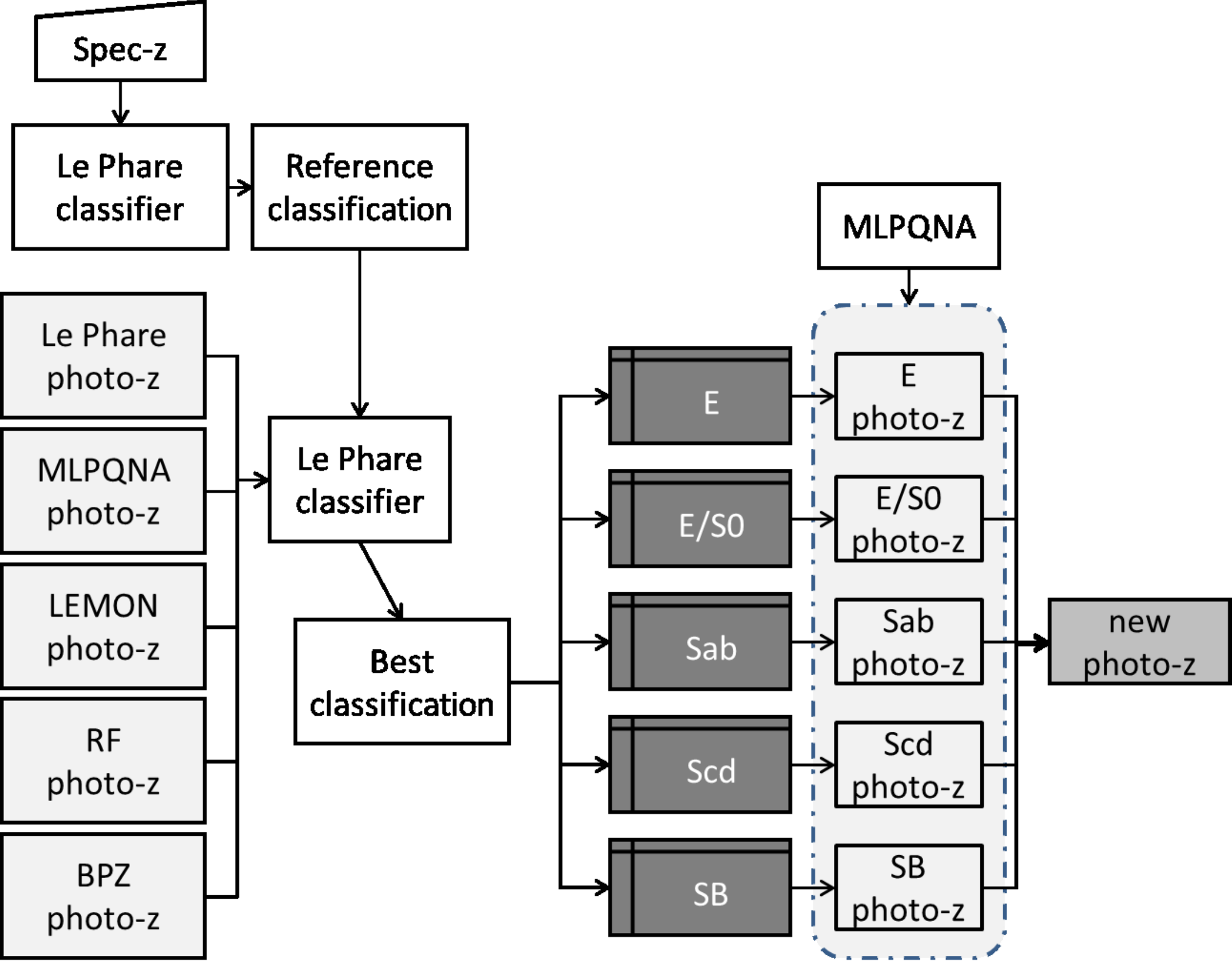}
\caption{Workflow of the method implemented to combine SED fitting and ML models to improve the overall photo-z estimation quality. See text for details.}\label{fig:workflow}
\end{figure}

The most relevant part of our work consisted in checking whether a combination of methods could be used to improve the overall results.  In order to investigate such possibility, we designed a hybrid approach, which makes use of both SED fitting and ML models, organized in a workflow structured in three main stages (Fig.~\ref{fig:workflow}).

\subsection{Preliminary experiments}\label{SEC:exp1}

First of all we tested the capability of each method to deal with data affected by different systematics, e.g. photometry not corrected for a) galactic extinction correction and/or b) the photometric zero-point offsets, as discussed in Sec.~\ref{SEC:thedata}. Four experiments were performed with each model:

\begin{itemize}
\item $EX_\text{clean}$ : full KB using the clean photometry corrected by galactic extinction and offset;
\item $EX_\text{ext}$ : full KB using the photometry corrected by galactic extinction only (i.e. affected by an offset);
\item $EX_\text{off}$ : full KB using the photometry corrected by offset only (i.e. affected by galactic extinction);
\item $EX_\text{no}$   : full KB using the photometry not corrected by offset and galactic extinction.
\end{itemize}

SED template fitting and empirical methods are differently affected by the dereddening (i.e. the correction for galactic extinction). In the first case, in fact, reddening introduces an artificial slope in the true SED, therefore, not taking it into account would affect photometric redshift estimates. In empirical methods, instead, since it affects in the same way also the objects in the training set, it should not affect the final outcome, at least if the parameter space is properly sampled.

We need to stress that even though fitting SED templates to magnitudes not corrected for the galactic extinction is not appropriate, the inclusion/exclusion of photometric offsets and dereddening helps to quantify how the redshifts derived with different methods are affected by the presence of systematics in the photometry.

Results are summarized in Tab.~\ref{TAB:STAT} for all the experiments. In Fig.~\ref{fig:scatters} we show the trends of the \zp\ vs. \zs\ for the test objects of the $EX_\text{clean}$ experiment using the $5$ considered models, where the MLPQNA model turned out to reach the best performance among all the explored methods. Fig.~\ref{fig:deltaz} displays the trends of $\Delta z$ vs. \zs\ for the same experiment and models.

\begin{table}
\centering
\begin{tabular}{ | l | l | l | l | l | l  }
\hline

	\textbf{EXP} & \textbf{MLPQNA} & \textbf{LEMON} & \textbf{RF} & \textbf{\LeP} & \textbf{BPZ}  \\ \hline\hline
	\multicolumn{6}{c}{bias}   \\ \hline\hline

	$EX_\text{clean}$ & 0.0007 & 0.0006 & 0.0010 & 0.0121 & 0.0289   \\
	$EX_\text{ext}$  & 0.0009 & 0.0009 & 0.0012  & 0.0183 & 0.0393   \\
	$EX_\text{off}$  & 0.0006 & 0.0007 & 0.0010  & 0.0158 & 0.0405    \\
	$EX_\text{no}$   & 0.0009 & 0.0010 & 0.0012  & 0.0225 & 0.0496    \\ \hline

	\multicolumn{6}{c}{$\sigma$}  \\ \hline\hline
	$EX_\text{clean}$ & 0.026 & 0.026 & 0.029 & 0.065 & 0.127  \\
	$EX_\text{ext}$  & 0.028 & 0.028 & 0.030  & 0.079 & 0.218  \\
	$EX_\text{off}$  & 0.026 & 0.026 & 0.029  & 0.066 & 0.142   \\
	$EX_\text{no}$   & 0.028 & 0.028 & 0.030  & 0.079 & 0.222   \\ \hline

	\multicolumn{6}{c}{$\sigma_{68}$} \\ \hline\hline
	$EX_\text{clean}$ & 0.018 & 0.018 & 0.021 & 0.041 & 0.039  \\
	$EX_\text{ext}$  & 0.021 & 0.020 & 0.023  & 0.048 & 0.039  \\
	$EX_\text{off}$  & 0.018 & 0.019 & 0.021  & 0.041 & 0.045  \\
	$EX_\text{no}$   & 0.021 & 0.020 & 0.023  & 0.049 & 0.043  \\ \hline

	\multicolumn{6}{c}{NMAD}  \\ \hline\hline
	$EX_\text{clean}$ & 0.018 & 0.018 & 0.021 & 0.038 & 0.031  \\
	$EX_\text{ext}$  & 0.020 & 0.020 & 0.022  & 0.044 & 0.034  \\
	$EX_\text{off}$  & 0.018 & 0.018 & 0.021  & 0.037 & 0.033  \\
	$EX_\text{no}$   & 0.020 & 0.020 & 0.022  & 0.044 & 0.034  \\ \hline

	\multicolumn{6}{c}{ \% Outliers}  \\ \hline\hline
	$EX_\text{clean}$ & 0.31 & 0.30 & 0.40 & 0.89 & 2.18  \\
	$EX_\text{ext}$  & 0.34 & 0.35 & 0.42  & 2.51 & 3.83  \\
	$EX_\text{off}$  & 0.31 & 0.29 & 0.39  & 1.12 & 3.21  \\
	$EX_\text{no}$   & 0.33 & 0.36 & 0.36  & 2.63 & 4.37  \\ \hline
\end{tabular}
\caption{Blind test set statistical results for the four experiment types with the five selected methods. The outlier percentage is reported according to the rule $|\Delta z/(z+1)|>0.15$.} \label{TAB:STAT}
\end{table}

\begin{table}
\centering
\begin{tabular}{|l|l|l|l|l|l|}
\hline
		   		&	 \textbf{MLPQNA}		&		\textbf{LEMON}		&		 \textbf{RF}		&		 \textbf{\LeP} 		&	  \textbf{BPZ}\\
\hline										
\hline										
			 \multicolumn{6}{c}{class \textit{E} - 2169 objects} \\\hline										
  $bias$ 		&		 -0.0007 		&		 -0.0004 		&		 0.0019 		&		 -0.0641 &		 -0.0297 		\\
  $\sigma$ 		&		 0.022	 		&		 0.022	 		&		 0.024	 		&		 0.045   &		 0.041 			\\
  $\sigma_{68}$ &		 0.016 			&		 0.016 			&		 0.017 			&		 0.086   &		 0.042 			\\
  $NMAD$ 		&		 0.015	 		&		 0.015	 		&		 0.016	 		&		 0.036   &		 0.027	 		\\
  $out. (\%)$ 	&		 0.18	 		&		 0.23	 		&		 0.28	 		&		 0.60    &		 0.65	 		\\
\hline										
			\multicolumn{6}{c}{class \textit{E/S0} - 1542 objects}\\ \hline										
  $bias$ 		&		 0.0001 		&		 -0.0002 		&		 -0.0035 		&		 0.0124  &		 -0.0381 		\\
  $\sigma$ 		&		 0.020	 		&		 0.019	 		&		 0.020	 		&		 0.029   &		 0.097 			\\
  $\sigma_{68}$ &		 0.014	 		&		 0.014	 		&		 0.016	 		&		 0.0267  &		 0.040 			\\
  $NMAD$ 		&		 0.014	 		&		 0.014	 		&		 0.015	 		&		 0.020   &		 0.024	 		\\
  $out. (\%)$ 	&		 0.26	 		&		 0.19	 		&		 0.2596 		&		 0.19    &		 3.11	 		\\
\hline										
			 \multicolumn{6}{c}{class \textit{Sab} - 1339 objects}\\ \hline										
  $bias$ 		&		 0.0007 		&		 0.0005 		&		 -0.0030 		&		 0.0073  &		 -0.0560 		\\
  $\sigma$ 		&		 0.024	 		&		 0.023	 		&		 0.026	 		&		 0.036   &		 0.186	 		\\
  $\sigma_{68}$ &		 0.019	 		&		 0.020	 		&		 0.023	 		&		 0.030   &		 0.050	 		\\
  $NMAD$ 		&		 0.019	 		&		 0.020	 		&		 0.022	 		&		 0.029   &		 0.034	 		\\
  $out. (\%)$ 	&		 0.07	 		&		 0.08	 		&		 0.15	 		&		 0.60    &		 5.23	 		\\
\hline										
			 \multicolumn{6}{c}{class \textit{Scd} - 3799 objects}\\ \hline										
  $bias$ 		&		 -0.0013 		&		 -0.0011 		&		 -0.0013 		&		 0.0022  &		 -0.0244 		\\
  $\sigma$ 		&		 0.026	 		&		 0.026	 		&		 0.031	 		&		 0.051   &		 0.112	 		\\
  $\sigma_{68}$ &		 0.020	 		&		 0.019	 		&		 0.023	 		&		 0.028   &		 0.036	 		\\
  $NMAD$ 		&		 0.019	 		&		 0.019	 		&		 0.023	 		&		 0.027   &		 0.031	 		\\
  $out. (\%)$ 	&		 0.32	 		&		 0.34	 		&		 0.47	 		&		 0.92    &		 1.61	 		\\
\hline										
  	 		 \multicolumn{6}{c}{class \textit{SB} - 1218 objects}\\ \hline										
  $bias$ 		&		 -0.0015 		&		 -0.0012 		&		 0.0003 		&		 -0.0163 &		 0.0005 		\\
  $\sigma$ 		&		 0.038	 		&		 0.036	 		&		 0.040	 		&		 0.121   &		 0.196	 		\\
  $\sigma_{68}$ &		 0.024	 		&		 0.023	 		&		 0.031	 		&		 0.043   &		 0.033	 		\\
  $NMAD$ 		&		 0.023	 		&		 0.023	 		&		 0.031	 		&		 0.041   &		 0.030	 		\\
  $out. (\%)$ 	&		 0.82	 		&		 0.66	 		&		 0.82	 		&		 2.55    &		 2.13	 		\\ \hline
\hline\end{tabular}										

\caption{Statistical results taken by considering the experiment $EX_\text{clean}$, by distinguishing among five spectral classes of galaxies, according to the original classification performed by \LeP\ (i.e. without bounding the fitting to any kind of redshift).}
\label{TAB:CLASSIFICATION}
\end{table}

\subsection{Classification based on template fitting}\label{SEC:standard}

The basic idea arose by analyzing the photo-z estimation results on the basis of the spectral-type classification, performed by \LeP\ without bounding the template fitting to any redshift estimate.
The statistical results summarized in Tab.~\ref{TAB:CLASSIFICATION}, show that the machine learning models provide a better performance for all spectral types. However, ML methods perform quite differently for the different spectral types individuated by \LeP. This induced us to explore the possibility to combine the methods: namely, the \LeP\ spectral-type classification is used to specialize ML methods and compute photo-z's for objects belonging to each spectral class. \\
Of course, the training of a specific regression model for each class can be effective only if the subdivision itself is as accurate as possible. A simple random subdivision could not enhance results. In fact, in the case of a random extraction of five subsets, the information contained in each single subset would be degraded, i.e. we would not gain any specialization but rather a reduction of patterns for each single regression network. Therefore, in this case, the best overall results would correspond to the precision achieved on  the whole dataset. Of course, it could happen that some subsets could improve the performance, but the overall results would be expected to remain either unchanged or get worse \citep{bishop}. Hence we needed the best subdivision, i.e. spectral-type classification, to proceed further.

After having obtained the $EX_\text{clean}$ results, we first defined the ``true'' spectral-type of each training galaxy as the best fitting spectral-type obtained with \LeP, constraining the redshift to its spectroscopic value.
We then used \LeP\ with the five different photometric redshift estimates, thus obtaining five different spectral-type predictions for each training galaxy.
The comparison of the ``true'' spectral types with the five different predictions shows that, in absence of spectroscopic information, Random Forest provides the most accurate spectral-type prediction.

The comparison among the different predictions is visualized (Fig.~\ref{fig:confmat}) by the  normalized confusion matrix \citep{provost1998}.
The confusion matrix is widely used to evaluate the performance of a classification: columns represent the instances in the predicted classes (the  classifier output) and rows give the expected (True) instances in the known classes. In a confusion matrix representing a two-class problem, displayed as an example in Tab.~\ref{conf_matrix}, the quantities are: $TP$ (True Positive), $TN$ (True Negative), $FP$ (False Positive) and $FN$ (False Negative). The example of confusion matrix in Tab.~\ref{conf_matrix} can be easily extended to the case with more than two classes: Fig.~\ref{fig:confmat} shows the case of $5$ spectral-type classes. Looking at the color bar close to each confusion matrix panel, reddish blocks contain higher percentages of objects, while the opposite occurs for bluish blocks. The ideal condition (i.e. the perfect classification for all classes) would correspond to have red all blocks on the main diagonal of the matrix and consequently in blue all other blocks. By comparing the five matrices, the RF model (panel c in Fig.~\ref{fig:confmat}) presents the best behavior for all classes.

\begin{table}
\centering
\begin{tabular}{| c | c | c | c |}
 & &\textbf{Predicted Labels}& \\
 &-&class 1&class 2\\
\hline
\hline
\textbf{True}&class 1&$TP$&$FN$\\
\textbf{Labels}&class 2&$FP$&$TN$\\
\hline
\end{tabular}
\caption{Structure of a confusion matrix for a two-class experiment. The interpretation of the symbols is self explanatory. For example, $TP$ denotes the number of objects belonging to the class $1$ that are correctly classified.}
\label{conf_matrix}
\end{table}

\subsection{Redshifts for spectral-type classes}\label{SEC:classes}

We then subdivided the KB on the base of the five spectral-type classes, thus obtaining five different subsets used to perform distinct training and blind test experiments, one for each individual class. The results for each class are depicted in figures \ref{fig:eclass}, \ref{fig:s0class}, \ref{fig:sabclass}, \ref{fig:scdclass}, \ref{fig:sbclass} and in Tab.~\ref{tab:class}. The figures confirm the statistical results of Tab.~\ref{tab:class}, where there is a clear improvement in the case of the combined approach for classes \textit{E}, \textit{E/S0} and \textit{Sab}, and all statistical estimators show better results than the \textit{standard} case. A similar behavior is visible for class \textit{Scd} with the only exception of bias, while in the case of \textit{SB} all estimators are better in the combined approach, with the exception of the $\sigma$ that remains unchanged.
The resulting amount of objects for each class is obviously different from the one displayed in Tab.~\ref{TAB:CLASSIFICATION}, which was based on a free fitting, i.e. with model template and redshift left free to vary.

\subsection{Recombination}\label{SEC:recombination}

The final stage of the workflow consisted into the recombination of the five subsets to produce the overall photo-z estimation, which was compared with the initial $EX_\text{clean}$ experiment in terms of the usual statistical performance. By considering the Tab.~\ref{tab:class}, the recombination statistics was calculated on the whole datasets, after having gathered together all the objects of all classes. The recombined results are reported in the last two rows of the Tab.~\ref{tab:class}. As already emphasized for single classes, all the statistical estimators show an improvement in the combined approach case, with the exception of a slightly worse bias.
Therefore, the statistics shown in Fig.~\ref{fig:allclass} and in Tab.~\ref{tab:class}, make it apparent that the proposed combined approach induces an estimation improvement for each class as well as for the whole dataset.

\begin{table}
\centering
\begin{tabular}{|l|r|r|r|r|r|r|r|}
\hline
  \textbf{$Class$} & \textbf{$Exptype$} & 	  \textbf{$Datasize$} &  \textbf{$bias$} &  \textbf{$\sigma$} &  \textbf{$NMAD$} &   \textbf{$out. (\%)$} & \textbf{$\sigma68$} \\
\hline
  \textbf{E} & \textbf{hybrid} 	& \textbf{638} 	& \textbf{-0.0009} & \textbf{0.020} & \textbf{0.016} & \textbf{0.00} & \textbf{0.017}\\
  E & standard 	& 638 	&  0.0130 & 0.029 & 0.022 & 0.31 & 0.028\\
  \textbf{E/S0} & \textbf{hybrid}	& \textbf{2858} 	& \textbf{-0.0005} & \textbf{0.016} & \textbf{0.012} & \textbf{0.10} & \textbf{0.012}\\
  E/S0 & standard 	& 2858 	& -0.0059 & 0.022 & 0.014 & 0.31 & 0.014\\
  \textbf{Sab} & \textbf{hybrid}	& \textbf{1383} 	& \textbf{-0.0003} & \textbf{0.015} & \textbf{0.015} & \textbf{0.00} & \textbf{0.014}\\
  Sab & standard 	& 1383 	& -0.0032 & 0.018 & 0.016 & 0.00 & 0.016\\
  \textbf{Scd} & \textbf{hybrid} 	& \textbf{3900} 	& \textbf{-0.0011} & \textbf{0.024} & \textbf{0.019} & \textbf{0.18} & \textbf{0.019}\\
  Scd & standard 	& 3900 	&  0.0006 & 0.025 & 0.020 & 0.23 & 0.020\\
  \textbf{SB} & \textbf{hybrid} 	& \textbf{1288} 	& \textbf{-0.0014} & \textbf{0.038} & \textbf{0.021} & \textbf{0.70} & \textbf{0.022}\\
  SB & standard 	& 1288 	&  0.0027 & 0.038 & 0.022 & 0.85 & 0.023\\\hline
  \textbf{ALL} & \textbf{hybrid} 	& \textbf{10067}	& \textbf{-0.0008} & \textbf{0.023} & \textbf{0.016} & \textbf{0.19} & \textbf{0.016}\\
  ALL & standard 	& 10067 & -0.0007 & 0.026 & 0.018 & 0.31 & 0.018\\
\hline
\end{tabular}
\caption{Photo-z estimation results, based on MLPQNA model, for each spectral-type subset of the test set, classified by \LeP\ by bounding the fit through the photo-z's predicted by RF model, which provided the best classification. The term \textit{hybrid} refers to the results obtained by the workflow discussed here and based on the combined approach, while \textit{standard} refers to the results obtained on the same objects but through the standard approach (i.e. $EX_\text{clean}$ experiment).}\label{tab:class}
\end{table}

\section{Discussion}\label{SEC:discussion}

As discussed in \cite{Cavuoti+15_KIDS_I} and confirmed in Tab.~\ref{TAB:STAT}, the MLPQNA outperforms SED fitting methods in all experiments. Instead, the other two empirical methods obtain results comparable with the MLPQNA. In particular, LEMON appears quite close to the MLPQNA in terms of results, a fact that could be expected by considering their similar learning laws \citep{shirangi2016}. We, however, preferred MLPQNA due to its better computational efficiency.

For the $EX_\text{clean}$ experiment we find a very small bias of $\sim 0.0007$, a standard deviation of $\sim 0.026$, a  $\sigma_{68}$ of $\sim 0.018$ a NMAD of $\sim 0.018$ and a number of outliers with $|\Delta z| > 0.15$ of only $0.31\%$ (see Tab.~\ref{TAB:STAT}). In contrast, the results from SED fitting methods are less accurate, with statistical estimators and outlier fractions worse than those found using ML methods. The presence of some objects scattered around $\zs \sim 0$ confirms that there is a small residual contamination from stars misclassified as galaxies.

Furthermore, by analyzing the statistics reported in Tab.~\ref{TAB:STAT}, it is evident that: \textit{i)} the presence of a photometric offset (experiment $EX_\text{ext}$) has a negligible impact on the performance of ML methods. In fact, almost all statistical estimators are the same as in the experiment with no corrections ($EX_\text{no}$); \textit{ii)} the results of ML methods are not affected by whether the input data are dereddened or not (experiment $EX_\text{off}$); \textit{iii)} \LeP\ is less affected by reddening than BPZ.
Therefore, the main contribution to the worse performance in the experiment $EX_\text{no}$ (without offset and reddening corrections) is due to the photometric offset.
On the contrary, the effects of a residual offset and reddening have a stronger impact on SED fitting methods, especially in terms of standard deviation and outliers fraction. The smaller impact on the $\sigma_{68}$ and NMAD estimators can be justified by considering their lower dependence on the presence of catastrophic outliers, which appears as the most relevant cause of a lower performance.

The spectral-type classification provided by the SED fitting method allows to derive also for ML models the statistical errors as function of spectral type, thus leading to a more accurate characterization of the errors. Therefore, it is possible to assign a specific spectral-type attribute to each object and to evaluate single class statistics. This fact, by itself, can be used to derive a better characterization of the errors. Furthermore, as it has been shown, the combination of SED fitting and ML methods allows also to build specialized (i.e. expert) regression models for each spectral-type class, thus refining the process of redshift estimation.

During the test campaign, we explored also the possibility to increase the estimation performance by injecting the photometric redshifts calculated with \LeP\ within the parameter space used for training. But the final statistical results were slightly worse of $\sim1\%$, revealing that at least in our case such parameter does not bring enough information.

Although the spec-z's are in principle the most accurate information available to bound the SED fitting techniques,
their use would make impossible to produce a reliable catalogue of photometric redshifts for objects not in the KB (
i.e. for objects not observed spectroscopically).
Thus, it appears reasonable to identify the best solution by making use of predicted photo-z's to bound fitting, in order to obtain a reliable spectral-type classification for the widest set of objects. This approach, having also the capability to use arbitrary ML and SED fitting methods, makes the proposed workflow widely usable in any survey project.

By looking at Tab.~\ref{tab:class}, our procedure shows clearly how the MLPQNA regression method benefits from the knowledge contribution provided by the combination of SED fitting (\LeP\ in this case) and machine learning (RF in the best case) classification stage. In fact, this allows to use a set of regression experts based on MLPQNA model, specialized to predict redshifts for objects belonging to specific spectral-type classes, thus gaining in terms of a better photo-z estimation.

By analyzing the results of Tab.~\ref{tab:class} in more detail, the improvement in photo-z quality is significant for all classes and for all statistical estimators, as also confirmed by the comparisons in the diagrams shown in figures \ref{fig:eclass}, \ref{fig:s0class},  \ref{fig:sabclass}, \ref{fig:scdclass} and \ref{fig:sbclass}. In fact, the diagrams of the residual distribution for classes \textit{E} and \textit{E/S0} show a better behavior for the combined approach in terms of distribution height and width. In the case of class \textit{Sab}, the residuals of the combined approach have a more peaked distribution. Only the two classes \textit{Scd} and \textit{SB} show a less evident improvement, since their residual distributions appear almost comparable in both experiment types, as confirmed by their very similar values of statistical parameters $\sigma$ and $\sigma_{68}$.
This leads to a more accurate photo-z prediction by considering the whole test set.

The only apparent exception is the mean (column \textit{bias} of Tab.~\ref{tab:class}), which suffers the effect of the alternation of positive and negative values in the \textit{hybrid} case that causes the algebraic sum to result slightly worse than the \textit{standard} case (the effect occurs on the fourth decimal digit, see column \textit{bias} of the last two rows of Tab.~\ref{tab:class}). This is not statistically relevant because the bias is one order of magnitude smaller than the $\sigma$ and  $\sigma_{68}$ and therefore negligible.

Special attention deserves the fact that in some cases, the \textit{hybrid} approach leads to the almost complete disappearance of catastrophic outliers. This is the case, for instance of the \textit{E} type galaxies. The reason is that for the elliptical galaxies the initial number of objects is lower than for the other spectral types in the KB. In the \textit{standard} case, i.e. the standard training/test of the whole dataset, such small amount of \textit{E} type representatives is mixed together with other more populated class objects, thus causing a lower capability of the method to learn their photometric/spectroscopic correlations. Instead, in the \textit{hybrid} case, using the proposed workflow, the possibility to learn \textit{E} type correlations through a regression expert increases the learning capabilities (see for instance Fig.~\ref{fig:confmat} and Fig.~\ref{fig:eclass}), thus improving the training performance and the resulting photo-z prediction accuracy.

In particular the confusion matrices shown in Fig.~\ref{fig:confmat} provide a direct visual impact and a quick comparison on the classification results.
Each confusion matrix shown is referred to the results of a different spectral-type classification performed by \LeP, by varying the photo-z's estimated through the five different regression models and used to bound the SED fitting procedure. Moreover, a confusion matrix allows also to compare classification statistics. The most important statistical estimators are: \textit{(i)} the \textit{purity} or \textit{precision}, defined as the ratio between the number of correctly classified objects of a class (the block on the main diagonal for that class) and the number of objects predicted in that class (the sum of all blocks of the column for that class); \textit{(ii)} the \textit{completeness} or \textit{recall}, defined as the ratio between the number of correctly classified objects in that class (the block on the main diagonal for that class) and the total number of (true) objects of that class originally present in the dataset (the sum of all blocks of the row for that class); \textit{(iii)} the \textit{contamination}, automatically defined as the reciprocal value of the \textit{purity}.

Of course, there is an obvious correspondence between the visualized color-level confusion matrix and the \textit{purity} and \textit{completeness} statistics of its classes. For example, from the visual analysis of Fig.~\ref{fig:confmat}, it is evident that \textit{Scd} and \textit{SB} spectral-type classes are well classified by all methods. {This is also confirmed by their statistics, since the \textit{purity} is, on average, around $88\%$ for \textit{Scd} and $87\%$ for \textit{SB}, with an averaged \textit{completeness} of, respectively, $91\%$ in the case of \textit{Scd} and $82\%$ for \textit{SB}.

Moreover, the confusion matrices show that the three classifications based on the machine learning models maintain a good performance in the case of \textit{E/S0} spectral-type class, reaching on average a \textit{purity} and a \textit{completeness} of $89\%$ for both estimators.

In the case of \textit{Sab} class, only the RF-based classification is able to reach a sufficient degree of efficiency ($78\%$ of \textit{purity} and $85\%$ of \textit{completeness}). In particular, for the two cases based on photo-z's predicted by SED fitting models, for the \textit{Sab} class the BPZ-based results are slightly more \textit{pure} than those based on \LeP\ ($68\%$ vs $66\%$) but much less \textit{complete} (49\% vs 63\%).

Finally, by analyzing the results on the \textit{E} spectral-type class, only the RF-based case is able to maintain a sufficient compromise between \textit{purity} ($77\%$) and \textit{completeness} ($63\%$). The classification based on \LeP\ photo-z's reaches a $69\%$ of completeness on the \textit{E} class, but shows an evident high level of contamination between \textit{E} and \textit{E/S0}, thus reducing its purity to the $19\%$. It must be also underlined that the intrinsic major difficulty to separate \textit{E} objects from \textit{E/S0} class is due to the partial co-presence of both spectral types in the class \textit{E/S0}, that may partially cause wrong evaluations by the classifier.

Furthermore, the fact that later types are less affected may be easily explained by considering that their templates are, on average, more homogeneous than for early-type objects.

All the above considerations lead to the clear conclusion that the classification performed by \LeP\ model and based on RF photo-z's achieves the best compromise between purity and completeness of all spectral-type classes. Therefore, its spectral classification has been taken as reference throughout the further steps of the workflow.

At the final stage of the proposed workflow, the photo-z quality improvements obtained by the expert MLPQNA regressors on single spectral types of objects induce a reduction of $\sigma$ from $0.026$ to $0.023$ and of $\sigma_{68}$ from $0.018$ to $0.016$ for the overall test set, besides the more relevant improvement for the E class ($\sigma$ from $0.029$ to $0.020$ and of $\sigma_{68}$ from $0.028$ to $0.017$). Such virtuous mechanism is mostly due to the reduction of catastrophic outliers. This significative result, together with the generality of the workflow in terms of choice of the classification/regression methods, demonstrates the possibility to optimize the accuracy of photo-z estimation through the collaborative combination of theoretical and empirical methods.

\section{Conclusions}\label{SEC:conclusions}

In this work we propose an original workflow designed to improve the photo-z estimation accuracy through a combined use of theoretical (SED fitting) and empirical (machine learning) methods.

The data sample used for the analysis was extracted from the ESO KiDS-DR2 photometric galaxy data, using a knowledge base derived from the SDSS and GAMA spectroscopic samples. The Kilo Degree Survey (KiDS) provides wide and deep galaxy datasets with a good image quality in the optical wavebands \textit{u, g, r} and \textit{i}.

For a catalogue of about $25,000$ galaxies with spectroscopic redshifts, we estimated photo-z's using five different methods: \textit{(i)} Random Forest; \textit{(ii)} MLPQNA (Multi Layer Perceptron with the Quasi Newton learning rule); \textit{(iii)} LEMON (Multi Layer Perceptron with the Levenberg-Marquardt learning rule); \textit{(iv)} \LeP\ SED fitting and \textit{(v)} the bayesian model BPZ. The results obtained with the MLPQNA on the complete KiDS-DR2 data have been discussed in \cite{Cavuoti+15_KIDS_I}, thus further details are provided there.

We find that, as also found in \cite{carrasco2014}, machine learning methods provide far better redshift estimates, with a lower scatter and a smaller number of outliers when compared with the results from SED fitting.  The latter, however, is able to provide very useful information on the galaxy spectral type. Such information can be effectively used to constrain the systematic errors and to better characterize potential catastrophic outliers. Furthermore, this classification can be used to specialize the training of regression machine learning models on specific types of objects. Throughout the application on KiDS data, by combining in a single collaborative framework both SED fitting and machine learning techniques, we demonstrated that the proposed workflow is capable to improve the photo-z prediction accuracy.

\section*{Acknowledgments}
The authors would like to thank the anonymous referee for extremely valuable comments and suggestions.
Based on data products from observations made with ESO Telescopes at the La Silla Paranal Observatory under programme IDs 177.A-3016, 177.A-3017 and 177.A-3018, and on data products produced by Target/OmegaCEN, INAF-OACN, INAF-OAPD and the KiDS production team, on behalf of the KiDS consortium. OmegaCEN and the KiDS production team acknowledge support by NOVA and NWO-M grants. Members of INAF-OAPD and INAF-OACN also acknowledge the
support from the Department of Physics \& Astronomy of the University of Padova, and of the Department of Physics of Univ. Federico II (Naples). CT is supported through an NWO-VICI grant (project number $639.043.308$). MB and SC acknowledge financial contribution from the agreement ASI/INAF I/023/12/1. MB acknowledges the PRIN-INAF 2014 {\it Glittering kaleidoscopes in the sky: the multifaceted nature and role of Galaxy Clusters}. GL acknowledges for partial funding from PRIN-MIUR 2011 \textit{The Dark Universe and the cosmic evolution of baryons: from present day surveys to Euclid}.

\bibliographystyle{mn2e}

\begin{thebibliography}{99}
\bibitem[\protect\citeauthoryear{Ahn et al.}{2012}]{ahn2012} Ahn, C.~P., Alexandroff, R., Allende Prieto, C., et al.\ 2012, ApJS, 203, 21
\bibitem[\protect\citeauthoryear{Albrecht et al.}{2006}]{Albrecht} Albrecht, A., Bernstein, G., Cahn, R., et al., 2006, arXiv:astro-ph/0609591
\bibitem[\protect\citeauthoryear{Annunziatella et al.}{2016}]{Annunziatella} Annunziatella, M., Mercurio, A., Biviano, A., et al., 2016, A\&A, 585, A160,
\bibitem[\protect\citeauthoryear{Arnouts et al.}{1999}]{arnouts1999} Arnouts, S., Cristiani, S., Moscardini, L., et al., 1999, MNRAS, 310, 540
\bibitem[\protect\citeauthoryear{Baum}{1962}]{Baum} Baum, W.~A., 1962, Proceedings from IAU Symposium, ed. G.C. McVittie, 15, 390
\bibitem[\protect\citeauthoryear{Beck et al.}{2016}]{beck2016} Beck, R., et al., 2016 MNRAS 460, 1371-1381
\bibitem[\protect\citeauthoryear{Benitez et al.}{2004}]{Benitez2004} Ben�tez, N., Ford, H., Bouwens, R., et al., 2004, ApJS, 150, 1, 1-18
\bibitem[\protect\citeauthoryear{Benitez}{2000}]{Benitez} Benitez, N., 2000, ApJ, 536, 571
\bibitem[\protect\citeauthoryear{Bertin \& Arnouts}{1996}]{bertin1996} Bertin, E., Arnouts, S., 1996, A\&AS, 117, 393
\bibitem[\protect\citeauthoryear{Bishop}{1995}]{bishop} Bishop, C. M., 1995, Neural Networks for Pattern Recognition. Oxford University Press.
\bibitem[\protect\citeauthoryear{Biviano et al.}{2013}]{biviano2013} Biviano, A., Rosati, P., Balestra, I., et al., 2013, A\&A, 558, A1, 22 pp.
\bibitem[\protect\citeauthoryear{Bolton et al.}{2012}]{bolton2012} Bolton, A.~S., Schlegel, D.~J., Aubourg, E., et al., 2012, AJ, 144, 144
\bibitem[\protect\citeauthoryear{Breiman}{2001}]{breiman2001} Breiman, L., 2001, Machine Learning, Springer Eds., 45, 1, 25-32
\bibitem[\protect\citeauthoryear{Brescia et al.}{2015}]{brescia2015} Brescia, M., Cavuoti, S., Longo, G., De Stefano, V., 2015, A\&A, 568, A126, 7 pp.
\bibitem[\protect\citeauthoryear{Brescia et al.}{2014}]{brescia2014} Brescia, M., Cavuoti, S., Longo, G., et al., 2014, PASP, 126, 942, 743-797
\bibitem[\protect\citeauthoryear{Brescia et al.}{2013}]{brescia2013} Brescia M., Cavuoti S., D'Abrusco R., Mercurio A., Longo G., 2013, ApJ, 772, 140
\bibitem[\protect\citeauthoryear{Bruzual \& Charlot}{2003}]{bruzual2003} Bruzual, G., Charlot, S., 2003, MNRAS, 344, 1000
\bibitem[\protect\citeauthoryear{Byrd et al.}{1994}]{byrd1994} Byrd, R.H., Nocedal, J., Schnabel, R.B., 1994, Mathematical Programming, 63, 4, pp. 129-156
\bibitem[\protect\citeauthoryear{Calzetti et al.}{2000}]{Calzetti+00} Calzetti, D., Armus, L., Bohlin, R.~C., et al., 2000, ApJ, 533, 682
\bibitem[\protect\citeauthoryear{Capozzi et al.}{2009}]{Capozzi} Capozzi, D., De Filippis, E., Paolillo, M., D'Abrusco, R., Longo, G., MNRAS, 396, 900 (2009)

\bibitem[\protect\citeauthoryear{Carrasco Kind \& Brunner}{2014}]{carrasco2014} Carrasco Kind, M. and {Brunner}, R.~J.,  MNRAS, 442, 3380-3399 (2014)

\bibitem[\protect\citeauthoryear{Cavuoti et al.}{2015a}]{Cavuoti+15_KIDS_I} Cavuoti, S., Brescia, M., Tortora, C., et al.\ 2015, MNRAS, 452, 3, 3100-3105
\bibitem[\protect\citeauthoryear{Cavuoti et al.}{2015b}]{cavuoti2015} Cavuoti, S., Brescia, M., De Stefano, V., Longo, G., 2015, Experimental Astronomy, Springer, Vol. 39, Issue 1, 45-71
\bibitem[\protect\citeauthoryear{Cavuoti et al.}{2014}]{cavuoti2014b} Cavuoti, S., Brescia, M., Longo, G., 2014, Proceedings of the IAU Symposium, Vol. 306, Cambridge University Press
\bibitem[\protect\citeauthoryear{Cavuoti et al.}{2012}]{cavuoti2012} Cavuoti, S., Brescia, M., Longo, G., Mercurio, A., 2012, A\&A, 546, 13
\bibitem[\protect\citeauthoryear{Chang \& Lin }{2011}]{chang2011} Chih-Chung Chang and Chih-Jen Lin, 2011, ACM Transactions on Intelligent Systems and Technology, 2, 27
\bibitem[\protect\citeauthoryear{Chen et al.}{2012}]{chen2012} Chen, Y.-M., Kauffmann, G., Tremonti, C.~A., et al., 2012, MNRAS, 421, 314
\bibitem[\protect\citeauthoryear{Coleman et al.}{1980}]{coleman1980} Coleman, G.~D., Wu, C.~-C., Weedman, D.~W., 1980, ApJS, 43, 393
\bibitem[\protect\citeauthoryear{Connolly et al.}{1995}]{Connolly} Connolly, A.~J., Csabai, I., Szalay, A.~S., et al., 1995, AJ, 110, 2655
\bibitem[\protect\citeauthoryear{Csabai et al.}{2003}]{Csabai} Csabai, I., Budavari, T., Connolly, A.~J., et al., 2003, AJ, 125, 580
\bibitem[\protect\citeauthoryear{de Jong et al.}{2015}]{deJong+15_KIDS_paperI} de Jong, J.~T.~A., Verdoes Kleijn, G.~A., Boxhoorn, D.~R., et al., 2015, A\&A, 582, A62, 26 pp.
\bibitem[\protect\citeauthoryear{Driver et al.}{2011}]{driver2011} Driver, S.~P., Hill, D.~T., Kelvin, L.~S., et al., 2011, MNRAS, 413, 971
\bibitem[\protect\citeauthoryear{Fotopoulou et al.}{2016}]{fotopoulou2016}Fotopoulou, S. et al. submitted to MNRAS
\bibitem[\protect\citeauthoryear{Hildebrandt et al.}{2008}]{Hildebrandt2008} Hildebrandt, H., Wolf, C., Benitez, N.,, 2008, A\&A, 480, 703
\bibitem[\protect\citeauthoryear{Hildebrandt et al.}{2010}]{Hildebrandt2010} Hildebrandt, H., Arnouts, S., Capak, P., et al., 2010, A\&A, 523, 31
\bibitem[\protect\citeauthoryear{Hogg et al.}{1998}]{Hogg} Hogg, D.~W., Cohen, J.~G., Blandford, R., Pahre, M.~A., 1998, ApJ, 115, 1418
\bibitem[\protect\citeauthoryear{Kim et al.}{2015}]{kim2015} Kim, E.~J., Brunner, R.~J.; Carrasco Kind, M., MNRAS, Volume 453, Issue 1, p.507-521 (2015
\bibitem[\protect\citeauthoryear{Kinney et al.}{1996}]{Kinney1996} Kinney, A.~L., Calzetti, D., Bohlin, R.C., 1996, AJ, 467, 38
\bibitem[\protect\citeauthoryear{Koo}{1985}]{Koo} Koo, D.~C., 1985, AJ, 90, 418
\bibitem[\protect\citeauthoryear{Koo}{1999}]{Koo2} Koo, D.~C., 1999, {\it Astronomical Society of the Pacific Conference Series, ed. Weymann, Storrie-Lombardi, Sawicki \& Brunner.}, Vol. 191, 3
\bibitem[\protect\citeauthoryear{Kuijken}{2011}]{Kuijken11} Kuijken, K., 2011, OmegaCAM: ESO's newest imager. ESO Messenger, 146, 8
\bibitem[\protect\citeauthoryear{Kuijken et al.}{2015}]{Kuijken+15_GL_KiDS} Kuijken, K., Heymans, C., Hildebrandt, H., et al., 2015, arXiv:1507.00738
\bibitem[\protect\citeauthoryear{Ilbert et al.}{2006}]{ilbert2006} Ilbert, O., Arnouts, S., McCracken, H.~J., et al., 2006, A\&A, 457, 841
\bibitem[\protect\citeauthoryear{Ilbert et al.}{2009}]{Ilbert+09} Ilbert, O., Capak, P., Salvato, M., et al., 2009, ApJ, 690, 1236
\bibitem[\protect\citeauthoryear{La Barbera et al.}{2008}]{labarbera2008} La Barbera, F., de Carvalho, R.~R., Kohl-Moreira, J.~L., et al., 2008, PASP, 120, 681
\bibitem[\protect\citeauthoryear{Liske et al.}{2015}]{liske2015} Liske, J., Baldry, I.~K., Driver, S.~P., et al., 2015, MNRAS, 452, 2, 2087-2126
\bibitem[\protect\citeauthoryear{Loh \& Spillar}{1986}]{Loh} Loh, E.~D., Spillar, E.~J., 1986, ApJ, 303, 154
\bibitem[\protect\citeauthoryear{Massarotti et al.}{2001a}]{aMassarotti} Massarotti, M., Iovino, A., Buzzoni, A., 2001a, A\&A, 368, 74
\bibitem[\protect\citeauthoryear{Massarotti et al.}{2001b}]{bMassarotti} Massarotti, M., Iovino, A., Buzzoni, A., Valls-Gabaud, D., 2001b, A\&A, 380, 425
\bibitem[\protect\citeauthoryear{Masters et al.}{2015}]{masters2015} Masters, D., Capak, P., Stern, D., et al., 2015, ApJ, 813, 1, 53
\bibitem[\protect\citeauthoryear{McCulloch \& Pitts}{1943}]{pitts1943} McCulloch, W., \& Pitts, W., 1943, Bulletin of Mathematical Biophysics 5 (4): 115-133
\bibitem[\protect\citeauthoryear{McFarland et al.}{2013}]{McFarland+13} McFarland, J.~P., Verdoes-Kleijn, G., Sikkema, G., et al., 2013, Experimental Astronomy, 35, 45
\bibitem[\protect\citeauthoryear{Napolitano et al.}{2015}]{Napolitano+15_proc_lensing_KiDS} Napolitano, N.~R., Covone, G., Roy, N., et al., 2015, arXiv:1507.00733
\bibitem[\protect\citeauthoryear{Nocedal and Wright}{2006}]{nocedal2006} Nocedal, J., Wright, S.~J., 2006, Numerical Optimization, 2nd Edition. Springer
\bibitem[\protect\citeauthoryear{Noll et al.}{2004}]{Noll} Noll, S., Mehlert, D., Appenzeller, I., et al., 2004, A\&A, 418, 885
\bibitem[\protect\citeauthoryear{Peacock et al.}{2006}]{Peacock} Peacock, J.~A., Schneider, P., Efstathiou, G., et al., 2006, {\it ESA-ESO Working Group on Fundamental Cosmology, Tech. Rep.}
\bibitem[\protect\citeauthoryear{Polletta et al.}{2007}]{Polletta+07} Polletta, M., Tajer, M., Maraschi, L., et al., 2007, Apj, 663, 81
\bibitem[\protect\citeauthoryear{Prevot et al.}{1984}]{Prevot+84} Prevot, M.~L., Lequeux, J., Prevot, L., Maurice, E., \& Rocca-Volmerange, B., 1984, A\&A, 132, 389
\bibitem[\protect\citeauthoryear{Provost et al.}{1998}]{provost1998} Provost, F., Fawcett, T., Kohavi, R., 1998, Proceedings of the 15th International Conference on Machine Learning. Morgan Kaufmann Eds., 445-553
\bibitem[\protect\citeauthoryear{Puschell et al.}{1982}]{Puschell} Puschell, J.~J., Owen, F.~N., Laing, R.~A., 1982, ApJ, 257, L57
\bibitem[\protect\citeauthoryear{Radovich et al.}{2015}]{Radovich+15_proc_clusters_KiDS} Radovich, M., Puddu, E., Bellagamba, F., et al., 2015, arXiv:1507.00743
\bibitem[\protect\citeauthoryear{Rosenblatt}{1961}]{rosenblatt1961}Rosenblatt F., 1961, Principles of Neurodynamics: Perceptrons and the Theory of Brain Mechanisms. Spartan Books, Washington DC
\bibitem[\protect\citeauthoryear{Russell \& Norvig}{2003}]{russell2003} Russell, S. \& Norvig, P., 2003, Artificial Intelligence: A Modern Approach, second edition, p. 733. Prentice Hall. ISBN 0-13-080302-2
\bibitem[\protect\citeauthoryear{Sanches et al}{2014}]{sanchez2014} Sanches et al. 2014 MNRAS, Volume 445, Issue 2, p.1482-1506 (2014)
\bibitem[\protect\citeauthoryear{Schlafly \& Finkbeiner}{2011}]{Schlafly_Finkbeiner11} Schlafly, E.~F., \& Finkbeiner, D.~P., 2011, ApJ, 737, 103
%\bibitem[\protect\citeauthoryear{Schlegel et al.}{1998}]{schlegel1998} Schlegel, D.~J., Finkbeiner, D.~P., Davis, M., 1998, ApJ, 500, 525
\bibitem[\protect\citeauthoryear{Shirangi \& Emerick}{2016}]{shirangi2016} Shirangi, M.~G., Emerick, A.~A., 2016, Journal of Petroleum Science and Engineering, Elsevier, Vol. 143, 258-271
\bibitem[\protect\citeauthoryear{Silva et al.}{1998}]{Silva+98} Silva, L., Granato, G.~L., Bressan, A., \& Danese, L., 1998, ApJ, 509, 103
\bibitem[\protect\citeauthoryear{Speagle \& Eisenstein}{2016}]{speagle2016} Speagle, J.S \& Eisenstein, D.J. submitted to MNRAS
\bibitem[\protect\citeauthoryear{Tortora et al.}{2016}]{Tortora+15_KiDS_compacts} Tortora, C., La Barbera, F., Napolitano, N.~R., et al., 2016, MNRAS, 457, 3, 2845-2854
\bibitem[\protect\citeauthoryear{Umetsu et al.}{2012}]{Keiichi} Umetsu, K., Medezinski, E., Nonino, M., et al., 2012, {\it ApJ}, 755, 1, 56
\bibitem[\protect\citeauthoryear{Wolpert}{1992}]{wolpert1992}Wolpert, D. H., Neural Networks, Volume 5, Issue 2, Pages 241-259 (1992)
\bibitem[\protect\citeauthoryear{Zitlau et al.}{2016}]{zitlau2016}Zitlau R. et al, accepted by MNRAS (2016)

\end{thebibliography}

%\begin{figure*}
%\centering
%\includegraphics[width=0.6\textwidth]{ex1_dz.pdf} (a)
%\includegraphics[width=0.6\textwidth]{ex6_dz.pdf} (b)
%\caption{Upper panel: Comparison of $\Delta z/(1+z)$ distributions in the range $]-0.15, +0.15[$ for the experiments done with the $5$ models.
%Panels are referred, respectively, to: (a) all data with the fully corrected photometry (experiment $EX_\text{clean}$), (b) all data with the uncorrected photometry (experiment $EX_\text{no}$).}
%\label{fig:DELTA2}
%\end{figure*}

%\begin{figure*}
%\centering
%\includegraphics[width=0.6\textwidth]{ex4_dz.pdf} (a)
%\includegraphics[width=0.6\textwidth]{ex5_dz.pdf} (b)
%\caption{Upper panel: Comparison of $\Delta z/(1+z)$ distributions in the range $]-0.15, +0.15[$ for the experiments done with the $5$ models.
%Panels are referred, respectively, to: (a) all data with the photometry corrected by galactic extinction only (experiment $EX_\text{ext}$), (b) all data with the photometry corrected by offset only (experiment $EX_\text{off}$).}
%\label{fig:DELTA3}
%\end{figure*}

\begin{figure*}
\centering
\includegraphics[width=0.45\textwidth]{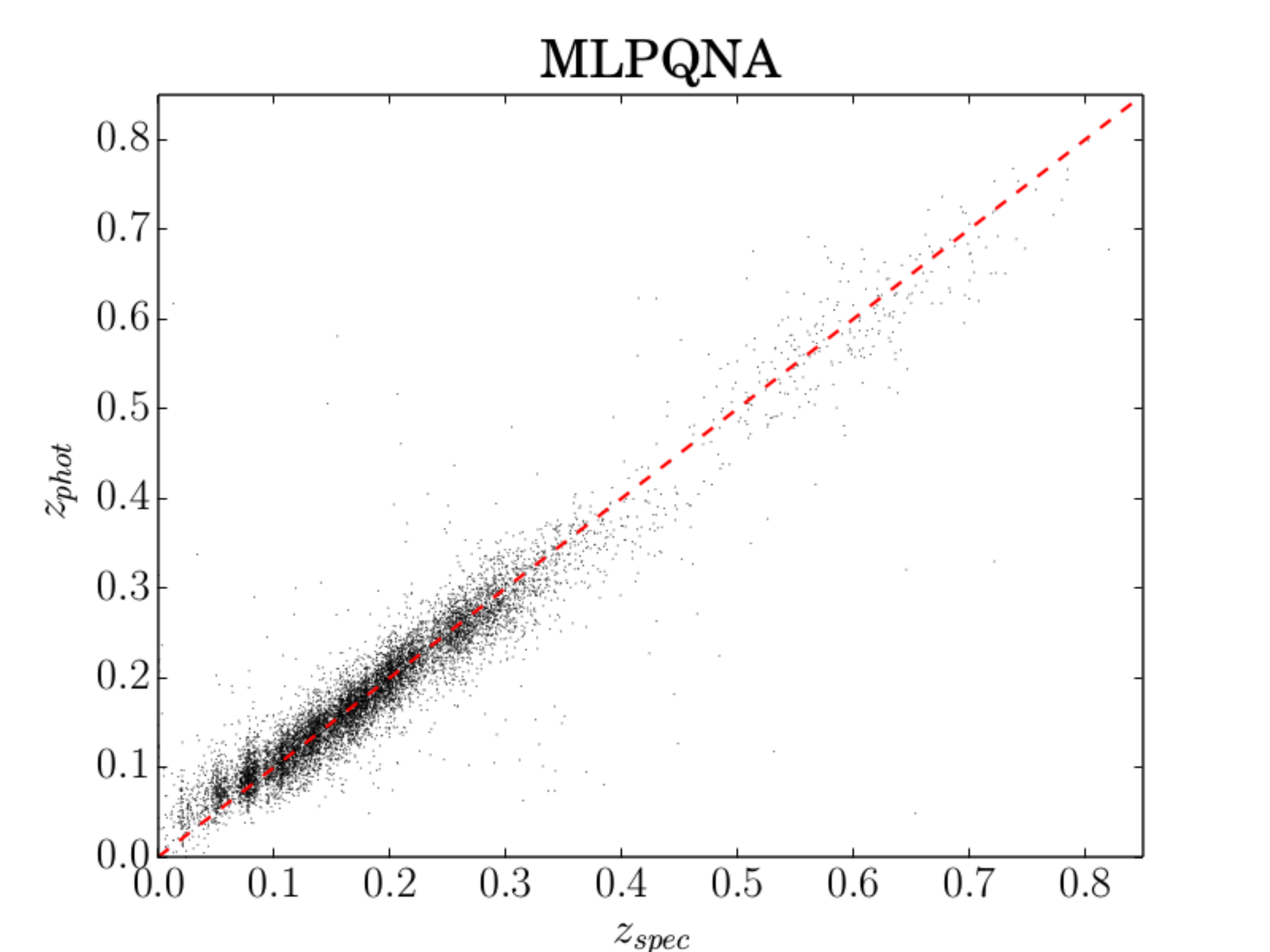} (a)
\includegraphics[width=0.45\textwidth]{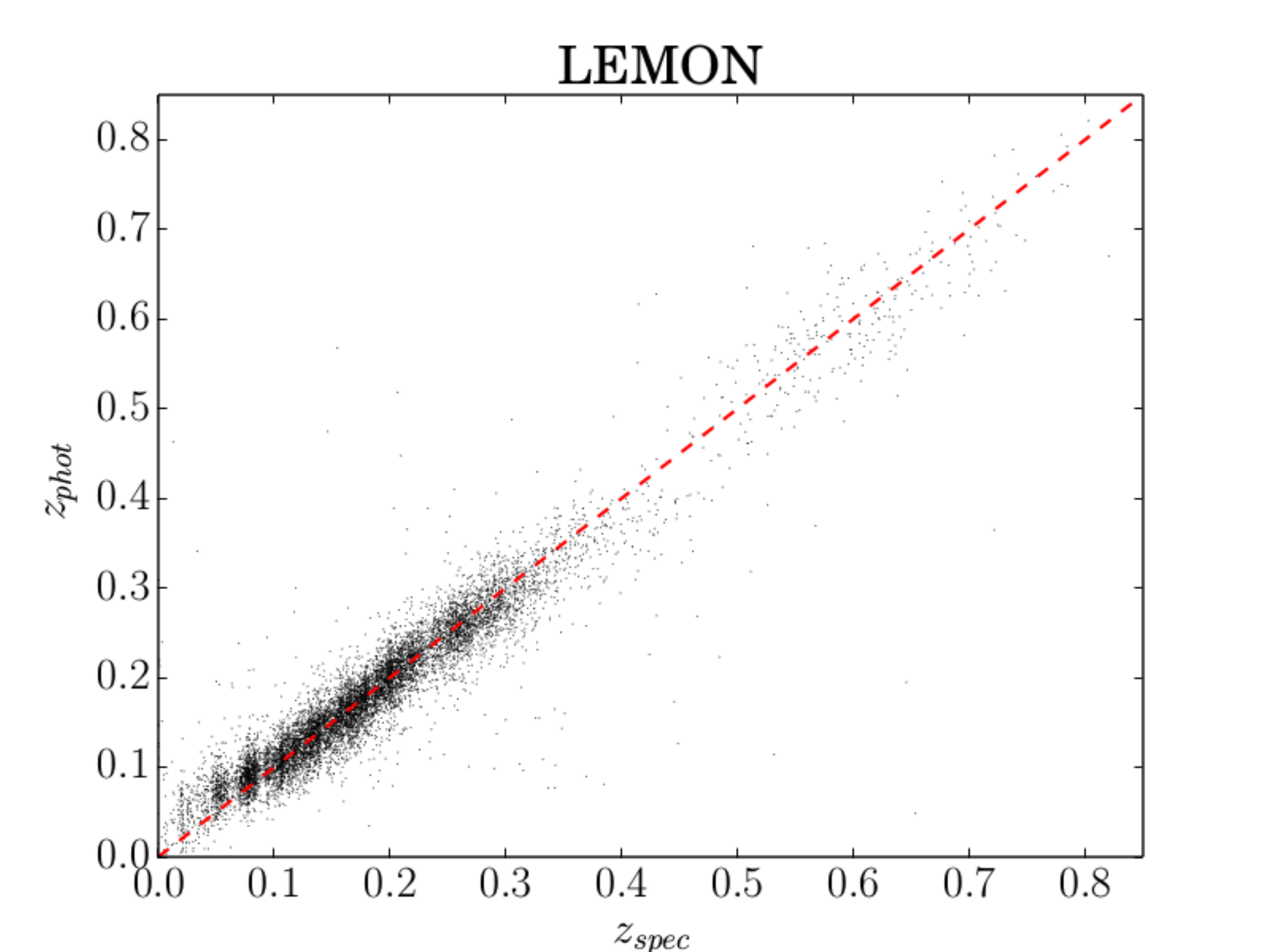} (b)
\includegraphics[width=0.45\textwidth]{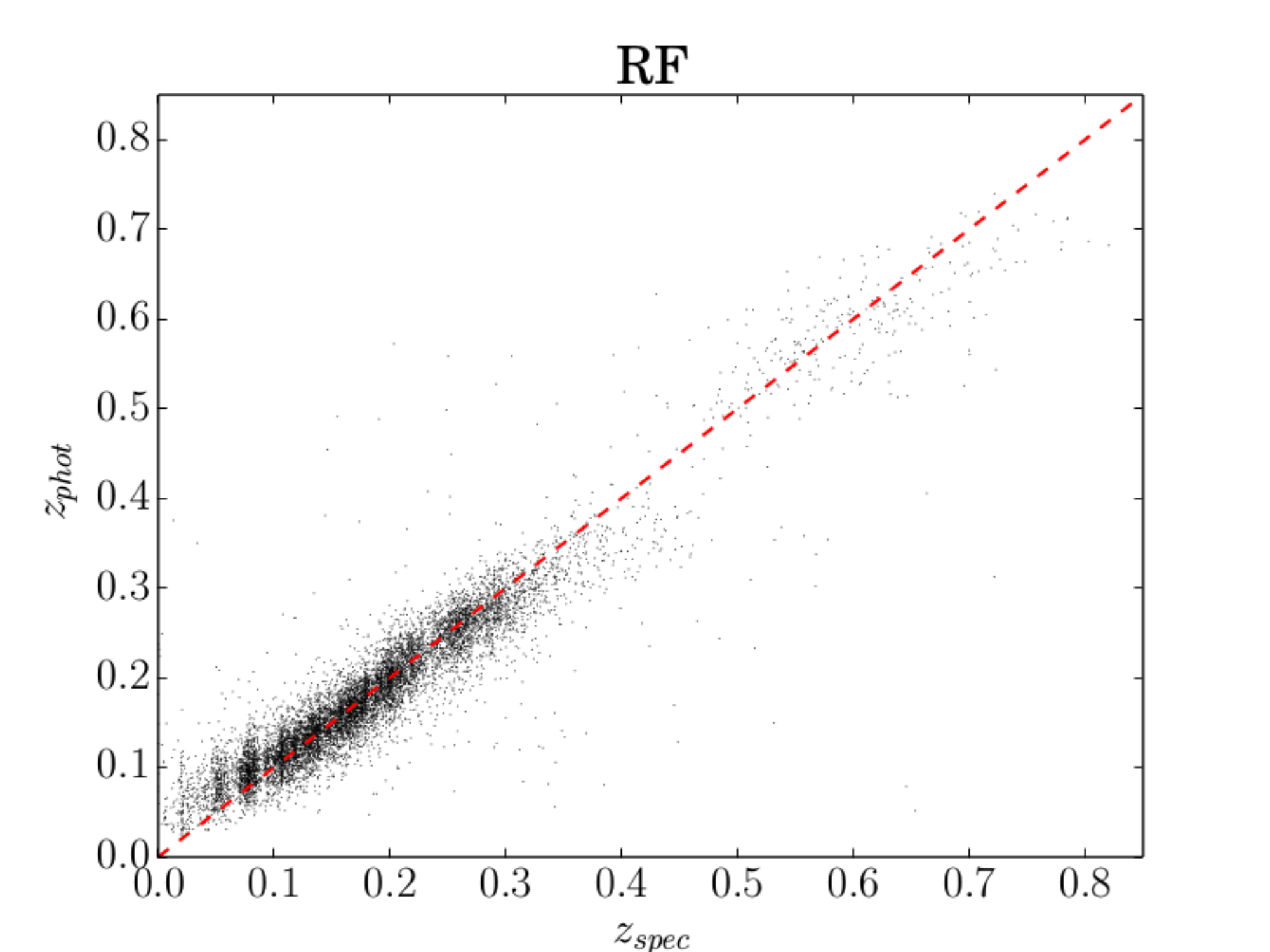} (c)
\includegraphics[width=0.45\textwidth]{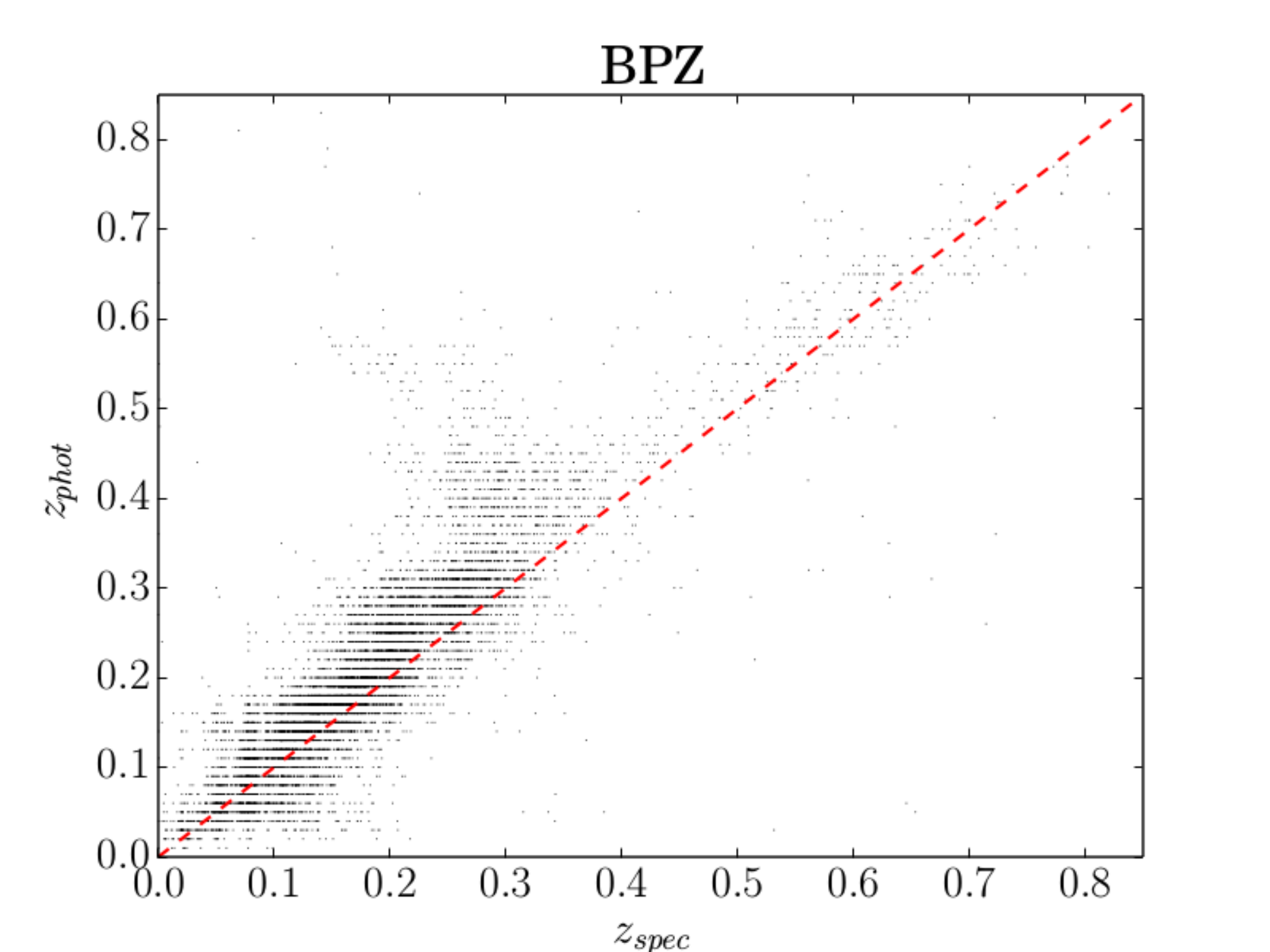} (d)
\includegraphics[width=0.45\textwidth]{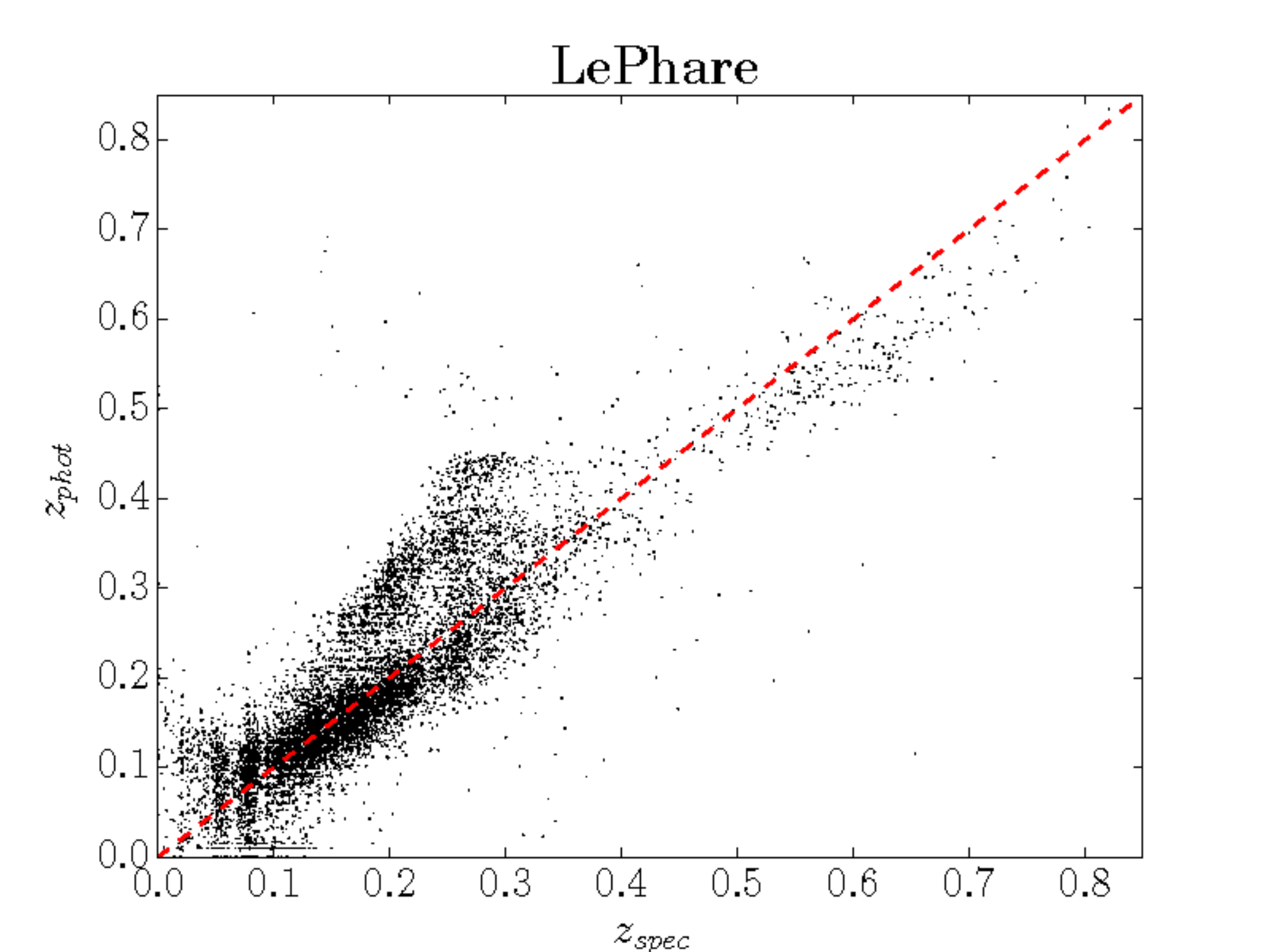} (e)
\caption{Diagrams of \zs\ vs. \zp\ for the data in the full redshift range available. Panels show results obtained in the case of the $EX_\text{clean}$ experiment by the various methods.}\label{fig:scatters}
\end{figure*}

\begin{figure*}
\centering
\includegraphics[width=0.45\textwidth]{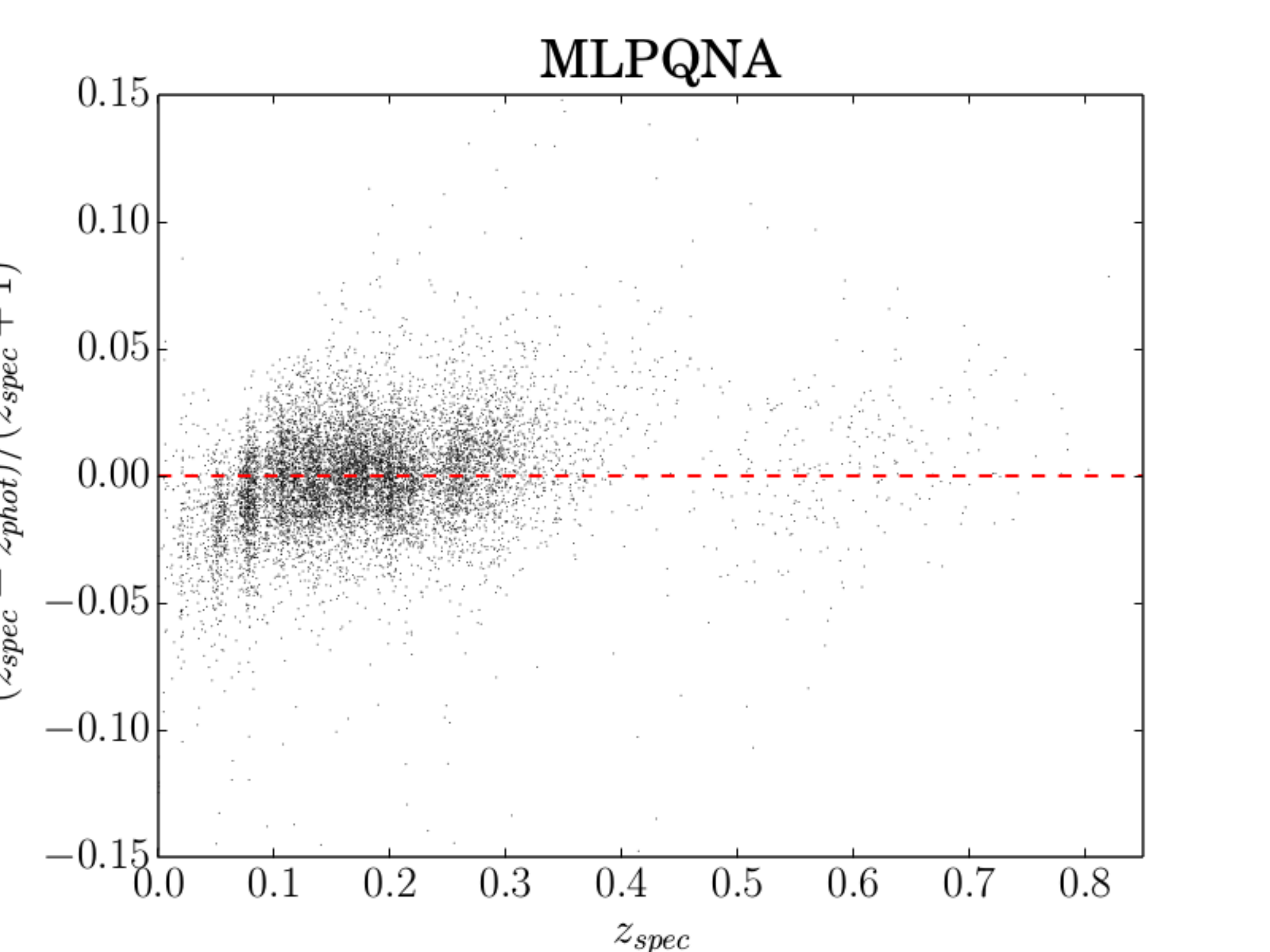} (a)
\includegraphics[width=0.45\textwidth]{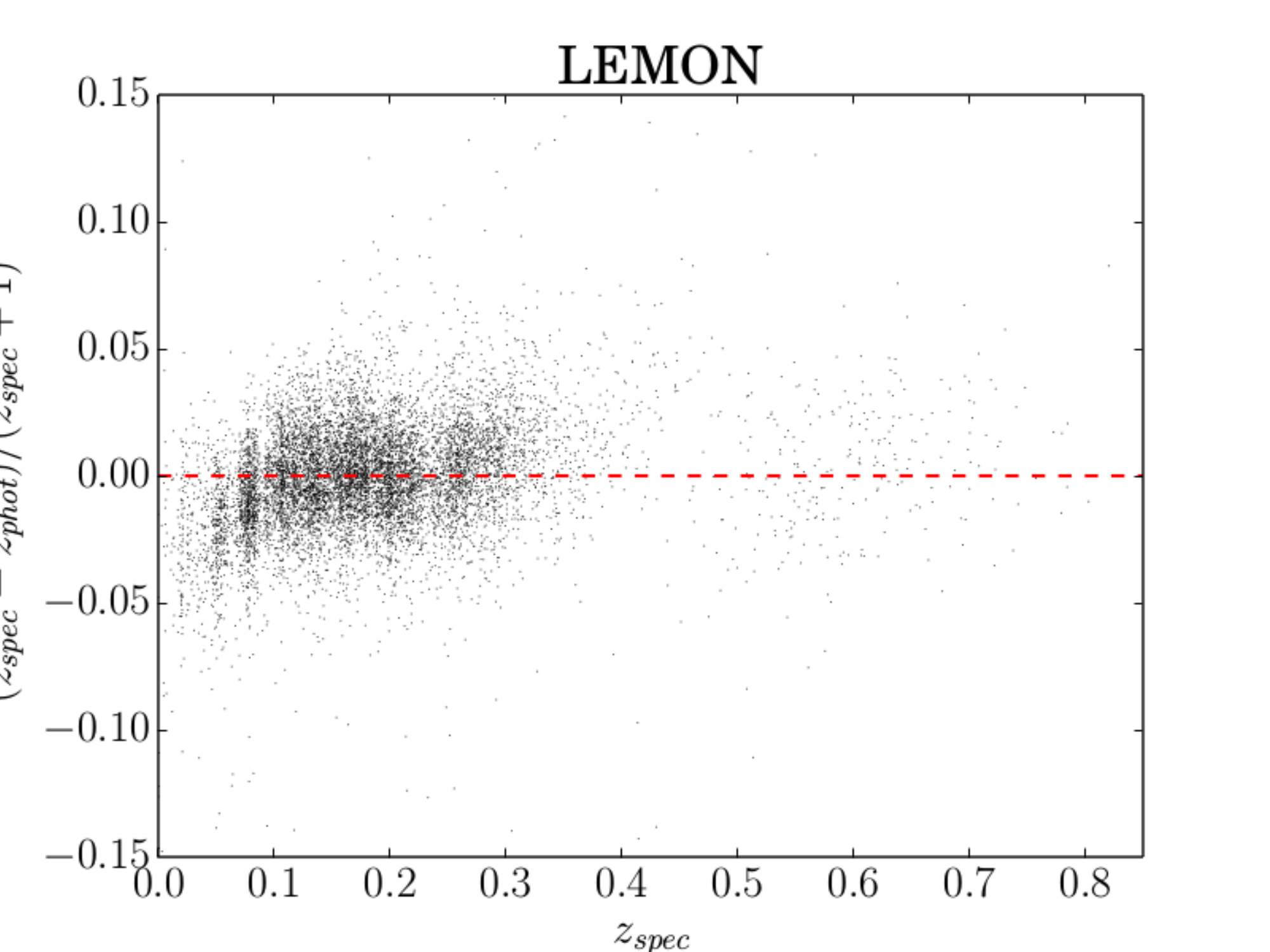} (b)
\includegraphics[width=0.45\textwidth]{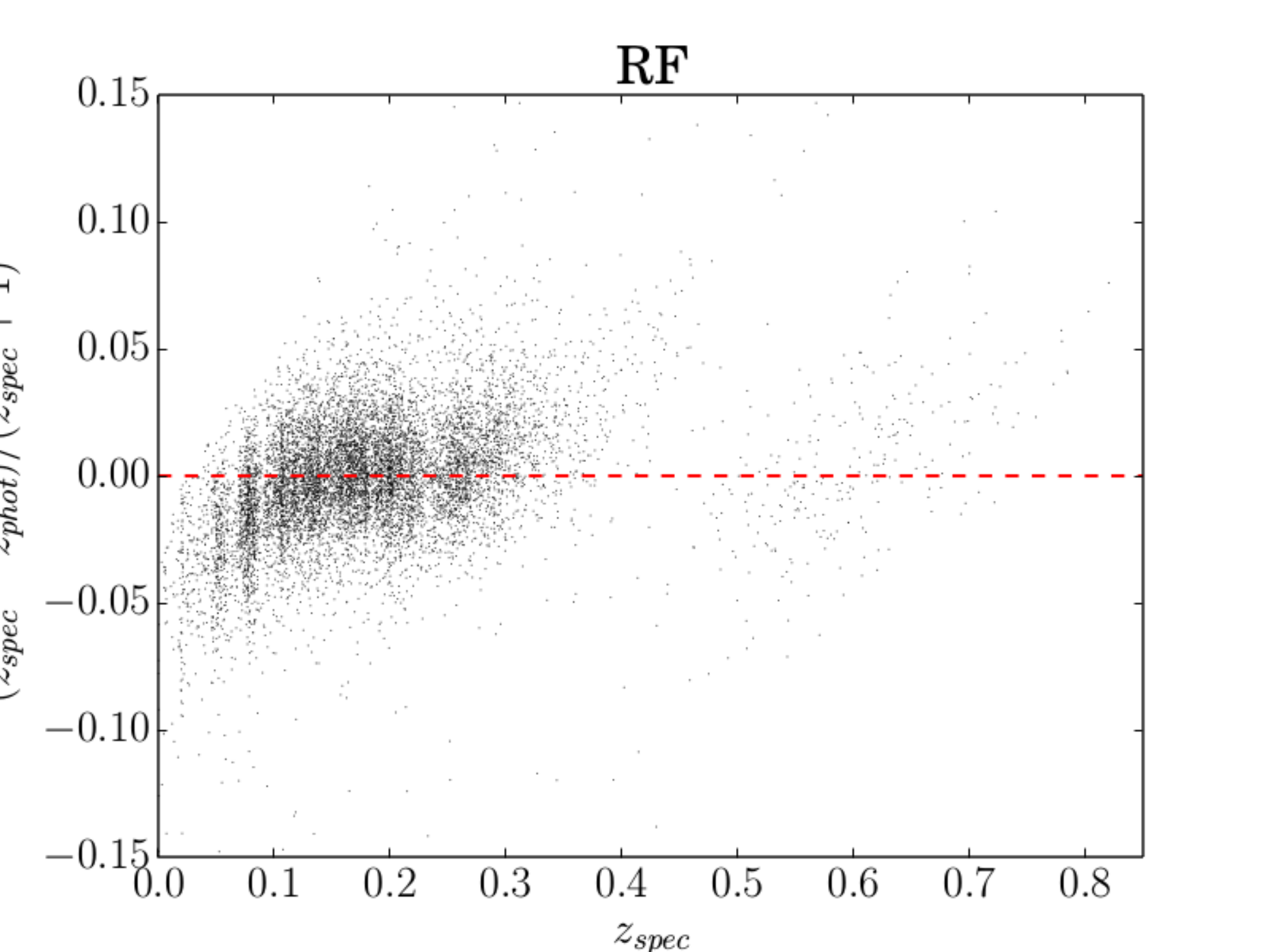} (c)
\includegraphics[width=0.45\textwidth]{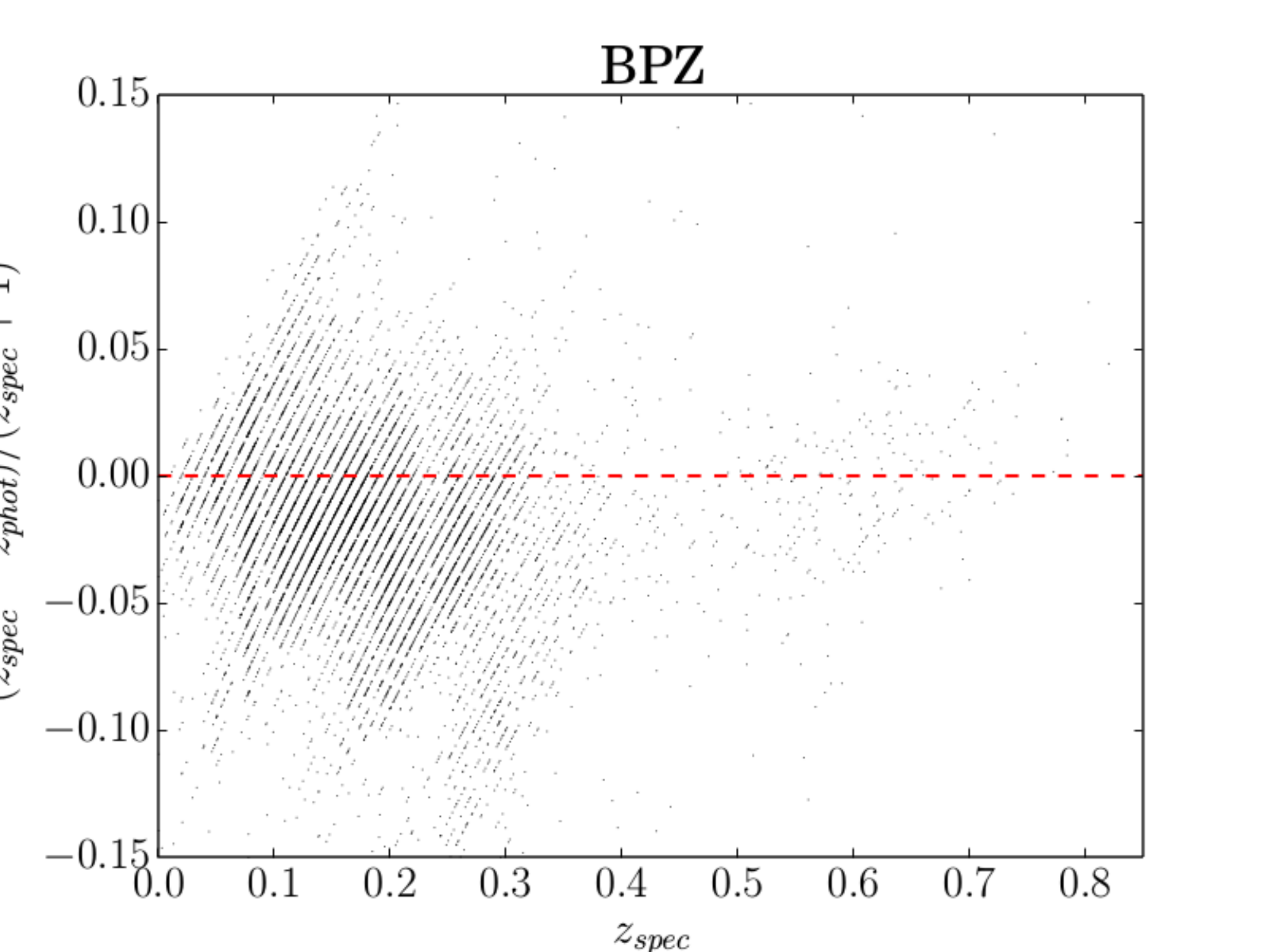} (d)
\includegraphics[width=0.45\textwidth]{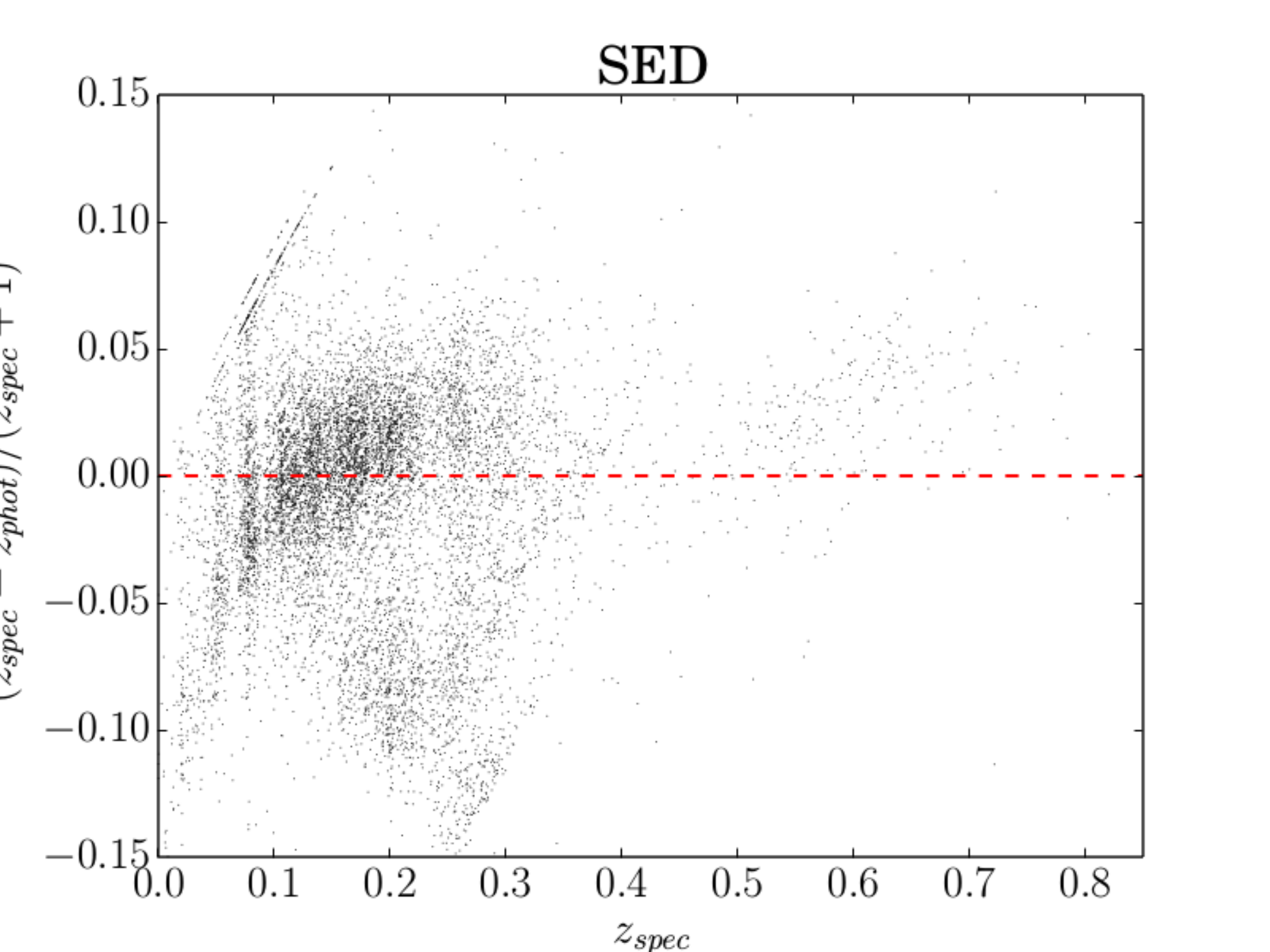} (e)
\caption{Diagrams of $\Delta z/(1+z)$ vs. $\zs$ diagrams for the  data in the full redshift range available. Panels show results obtained in the case of the $EX_\text{clean}$ experiment by the various methods.}\label{fig:deltaz}
\end{figure*}

\begin{figure*}
\centering
\includegraphics[width=0.45\textwidth]{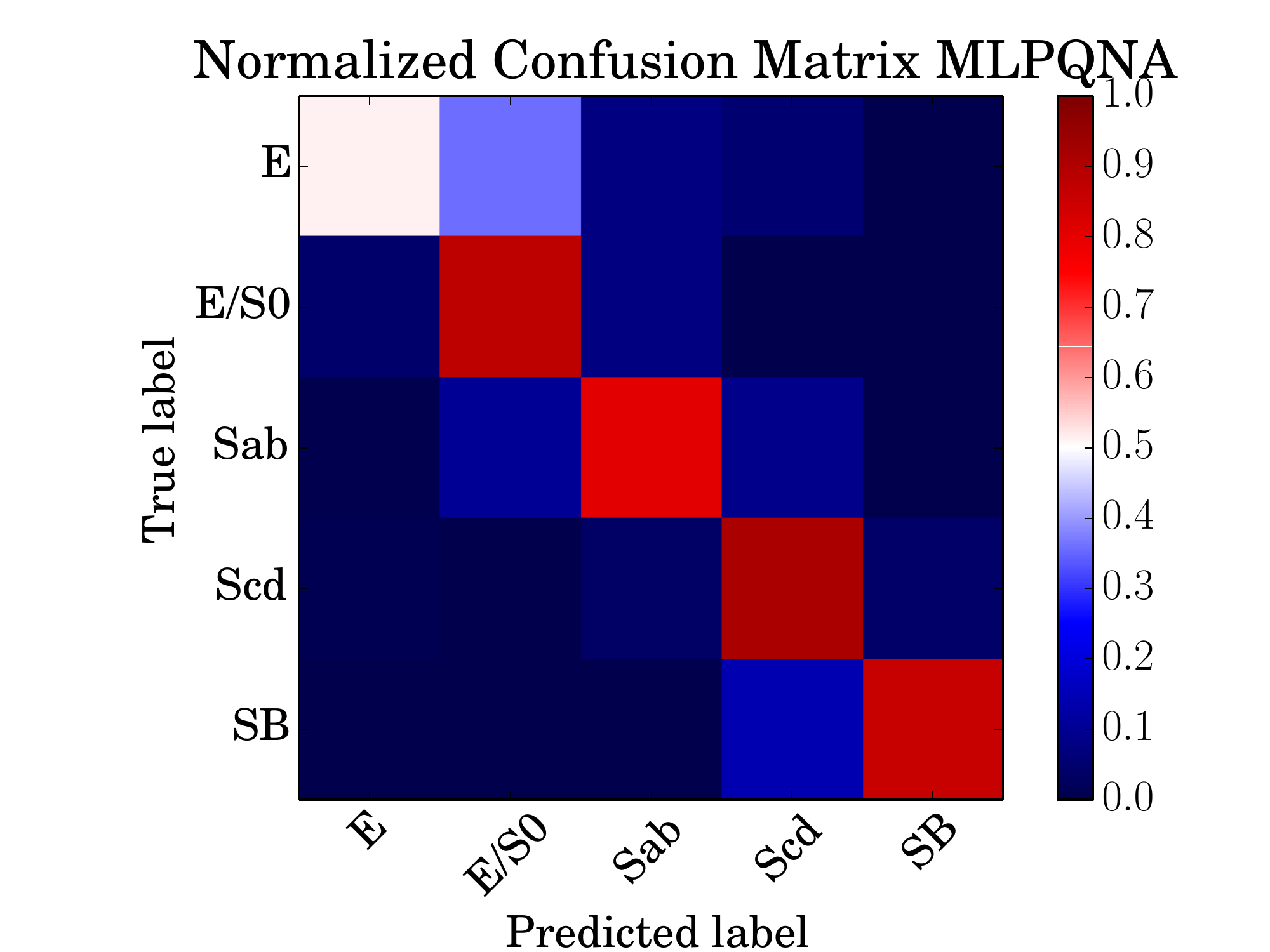} (a)
\includegraphics[width=0.45\textwidth]{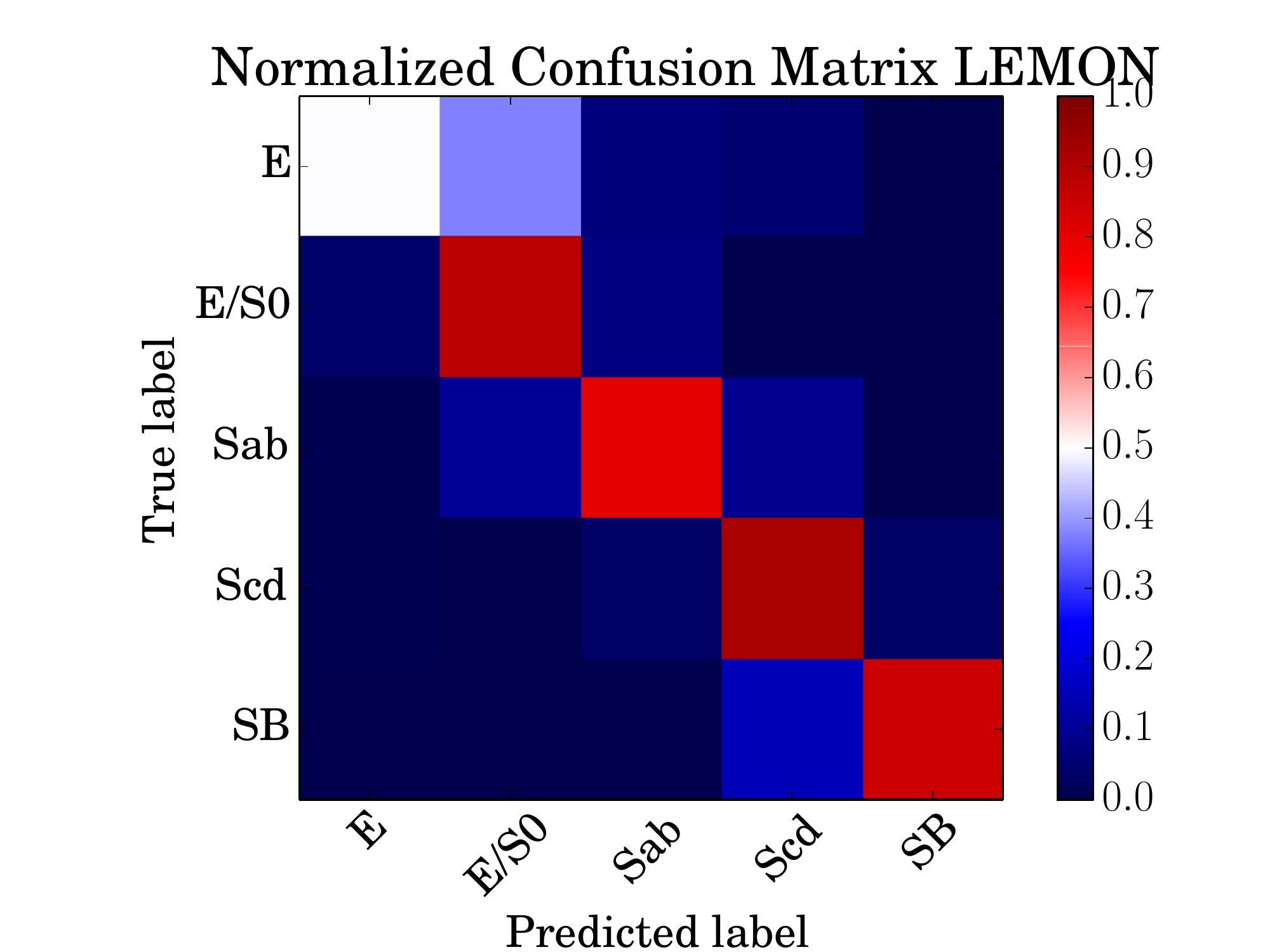} (b)
\includegraphics[width=0.45\textwidth]{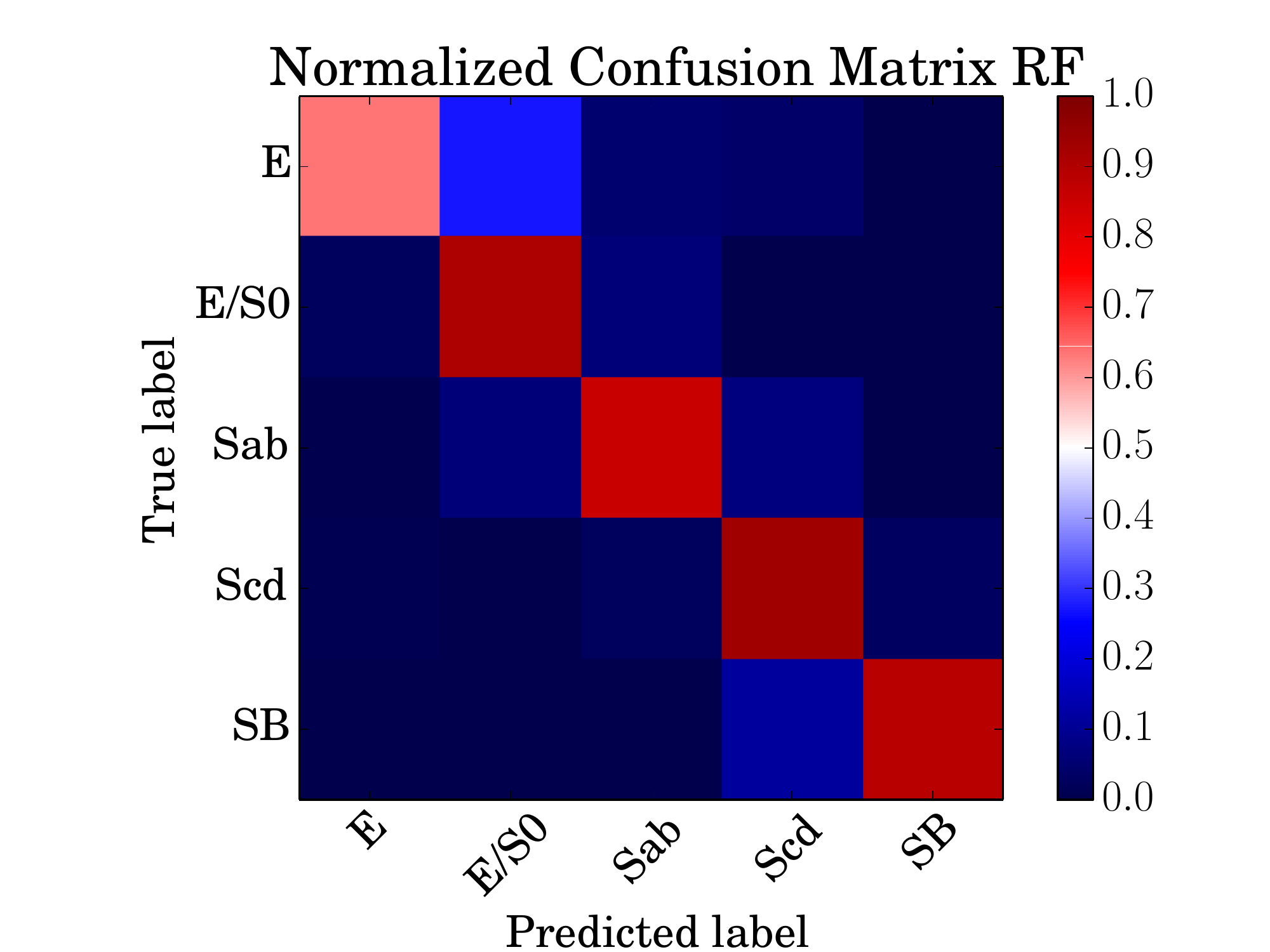} (c)
\includegraphics[width=0.45\textwidth]{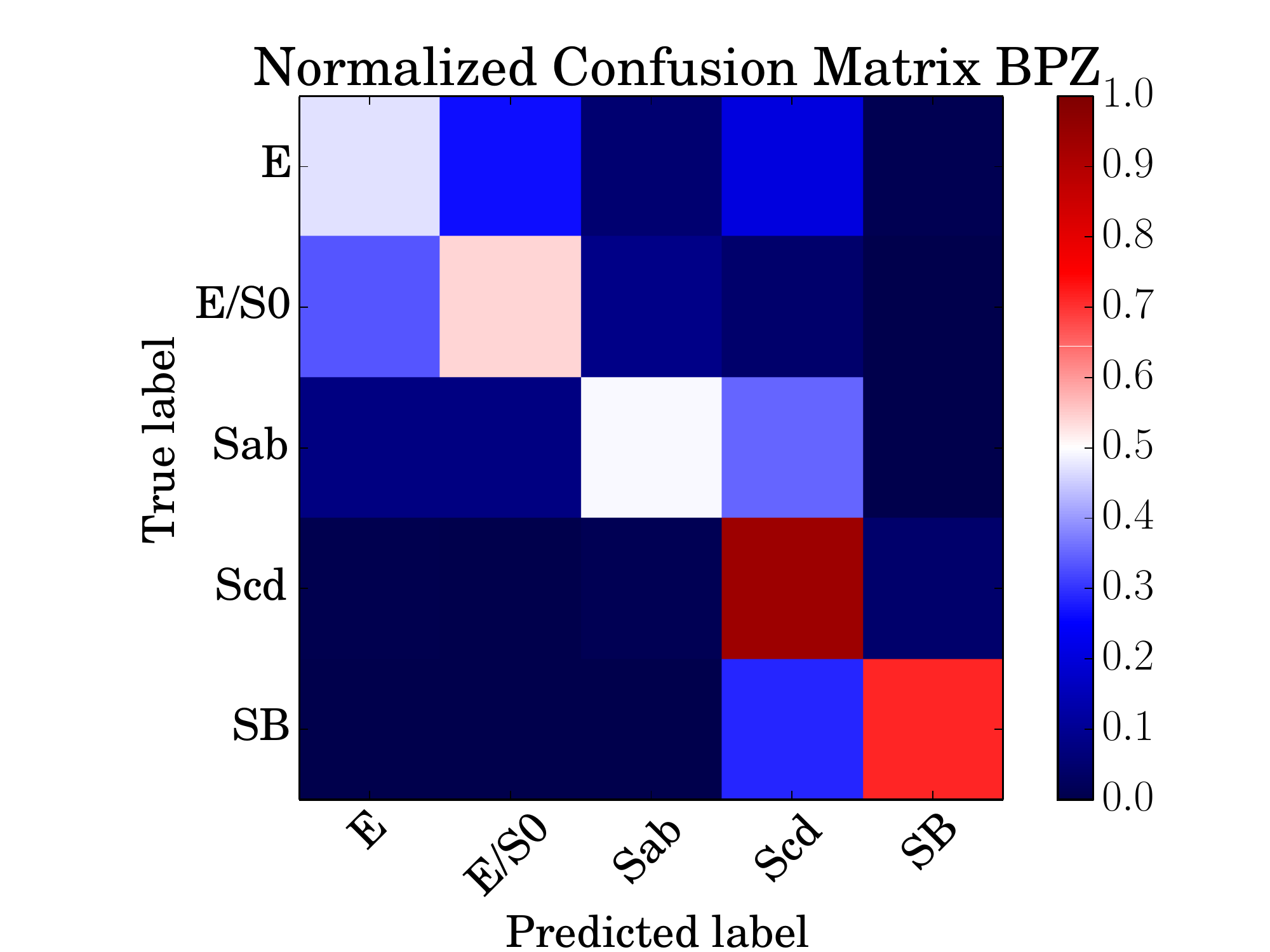} (d)
\includegraphics[width=0.45\textwidth]{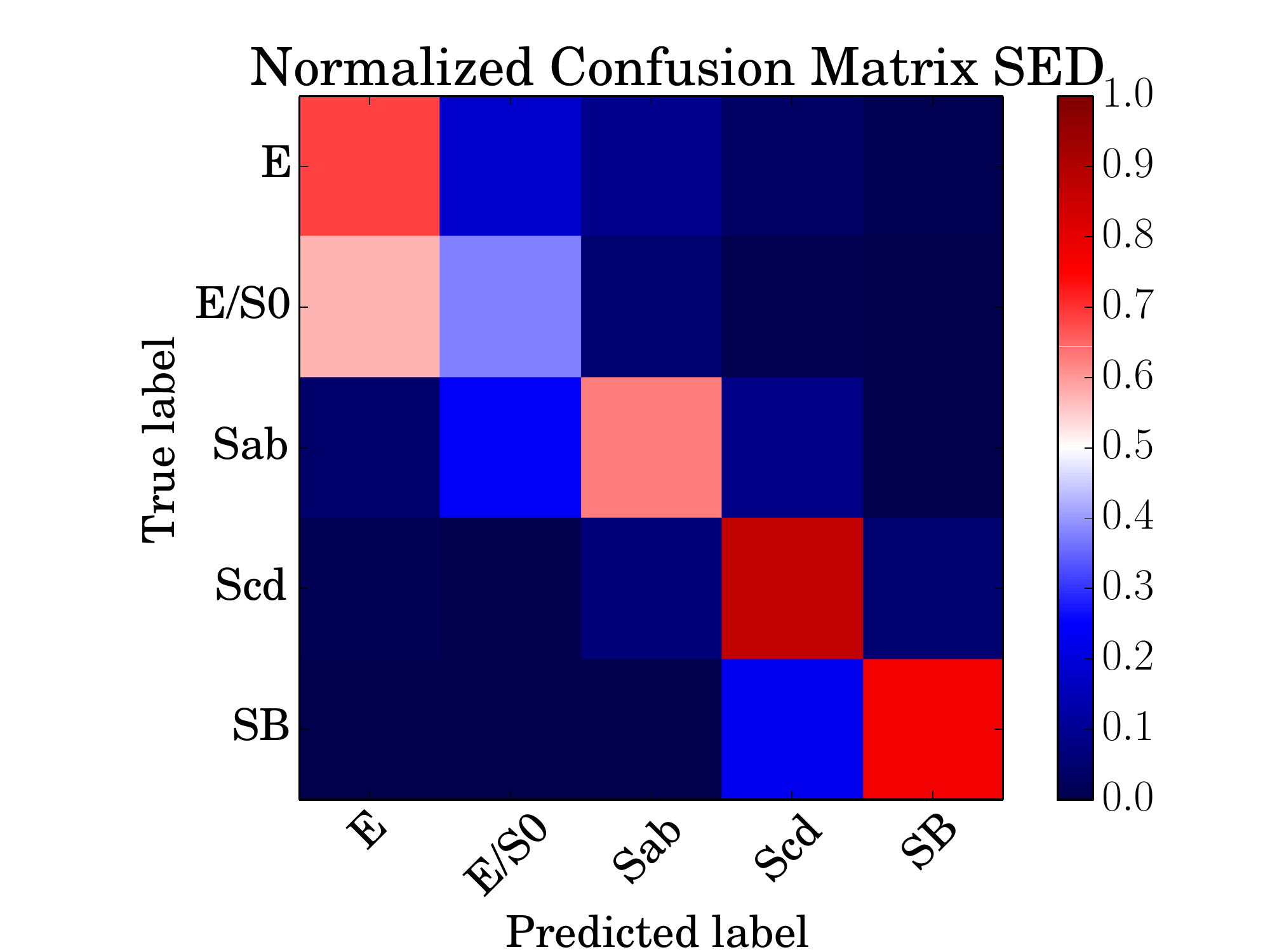} (e)
\caption{Normalized Confusion Matrices. The panels show \LeP\ classification results obtained by bounding the fitting with photo-z's derived, respectively, by (a) MLPQNA, (b) LEMON, (c) RF, (d) BPZ, (e) \LeP\ models, based on the $EX_\text{clean}$ experiment type. Reddish blocks include higher percentages of objects, while the opposite occurs for bluish blocks. The ideal condition (perfect classification for all classes) would correspond to have red all blocks on the main diagonal of the matrix and consequently in blue all other blocks.}\label{fig:confmat}
\end{figure*}

\begin{figure*}
\centering
\includegraphics[width=0.45\textwidth]{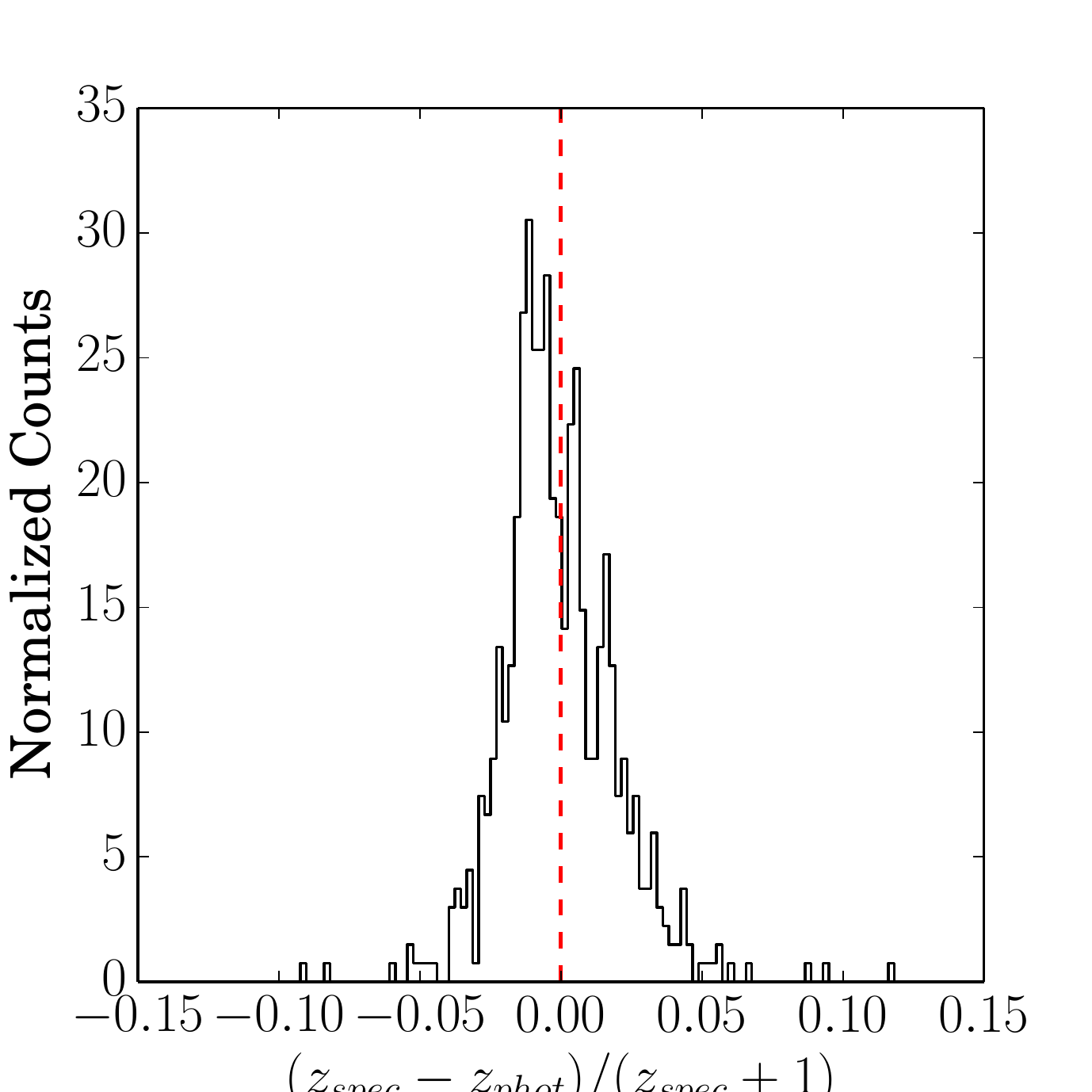} (a)
\includegraphics[width=0.45\textwidth]{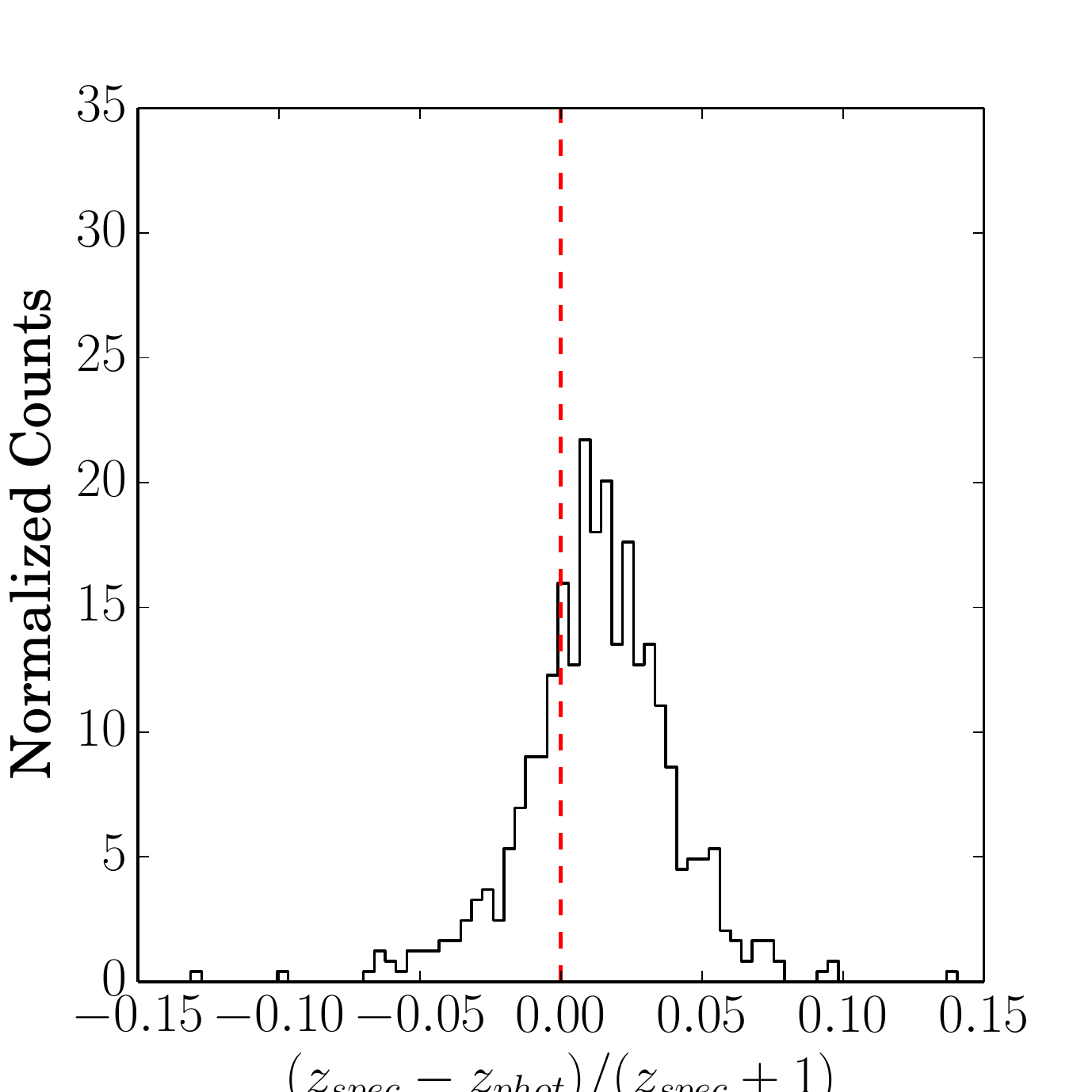} (b)
\caption{Histograms of $\Delta z/(1+z)$ in the case of \textit{E} class: left panel represents the results obtained by the \emph{expert} MLPQNA regressor through the proposed workflow, while the right panel represents the results obtained by the standard MLPQNA for the same objects.}\label{fig:eclass}
\end{figure*}

\begin{figure*}
\centering
\includegraphics[width=0.45\textwidth]{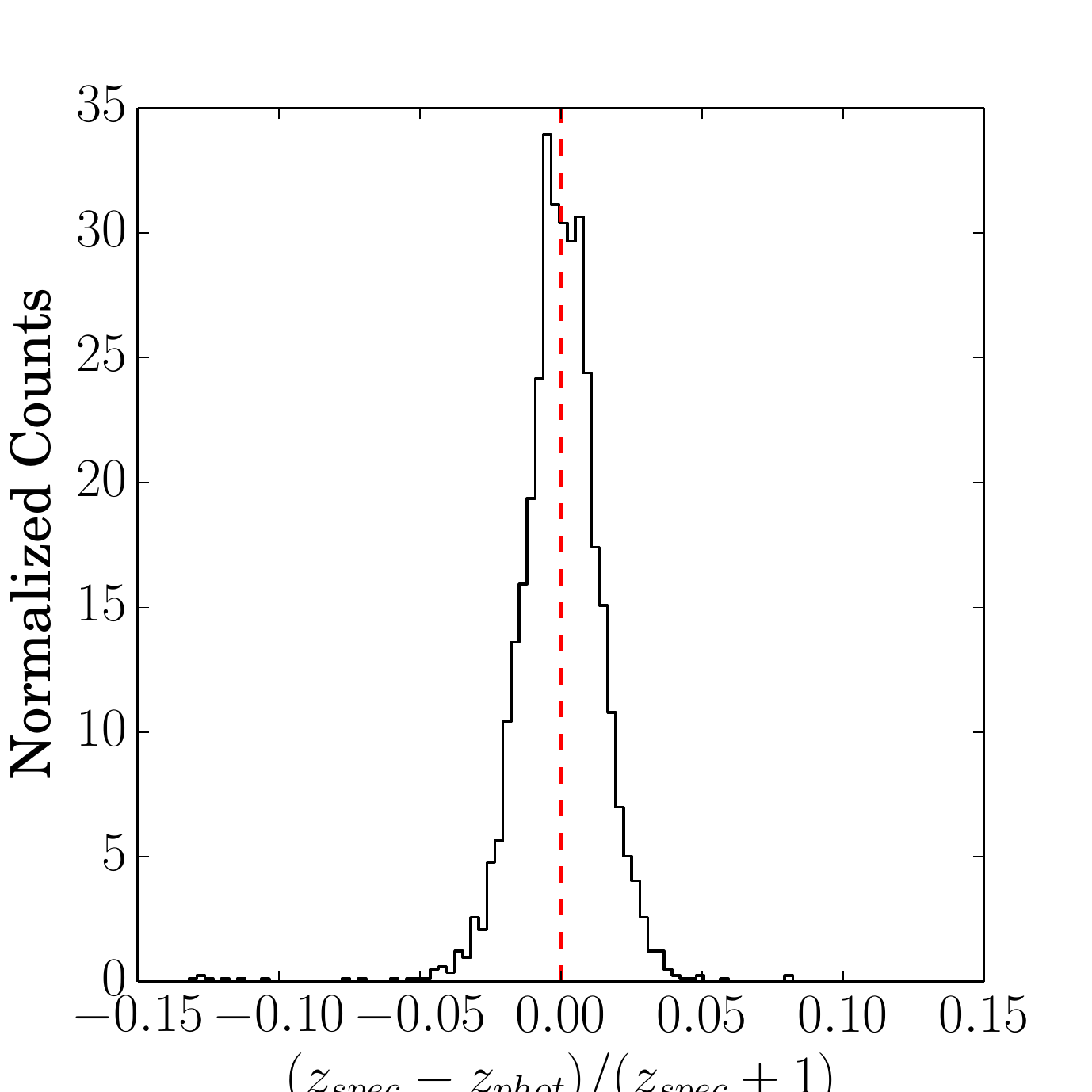} (a)
\includegraphics[width=0.45\textwidth]{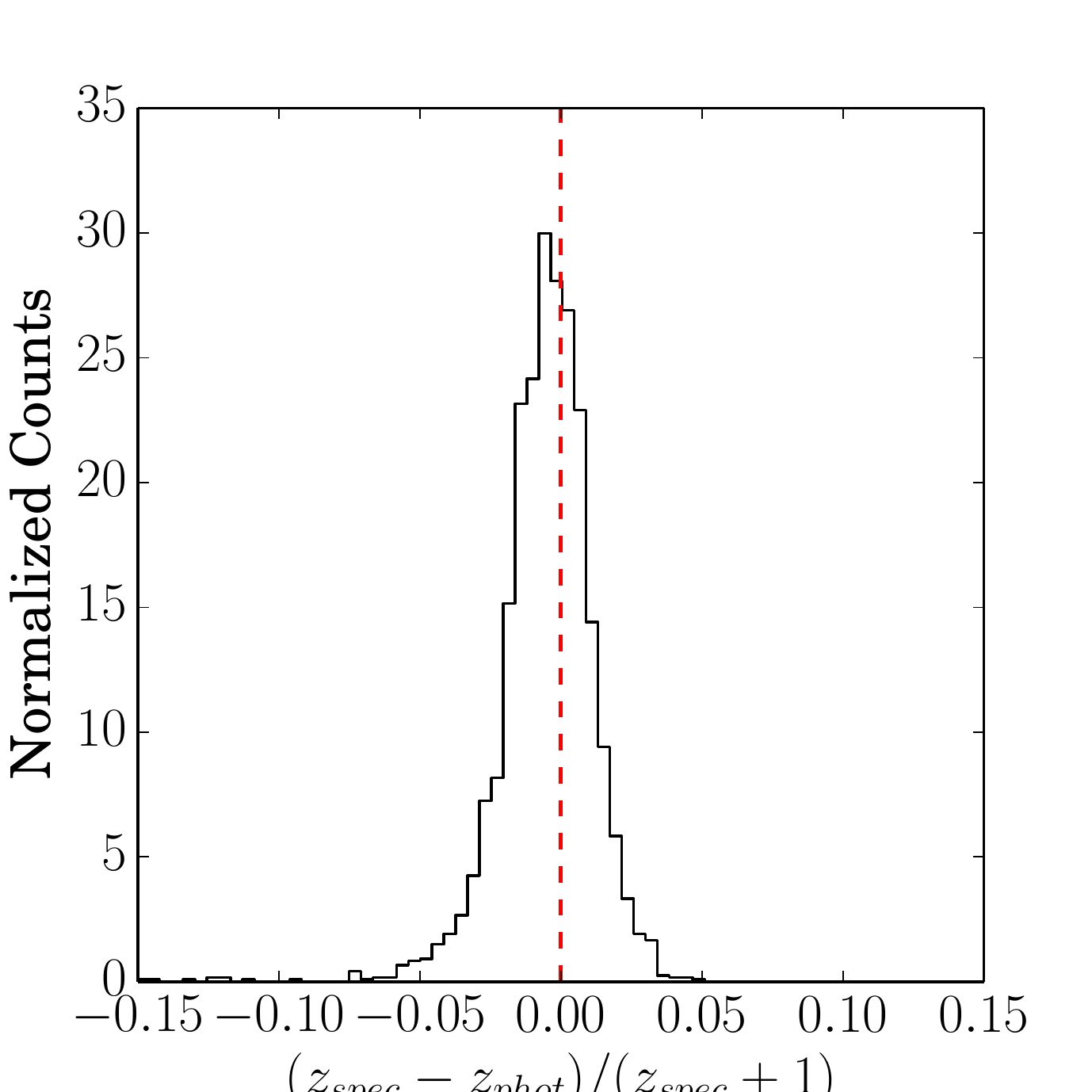} (b)
\caption{Histograms of $\Delta z/(1+z)$ in the case of \textit{E/S0} class: left panel represents the results obtained by the \emph{expert} MLPQNA regressor through the proposed workflow, while the right panel represents the result obtained by the standard MLPQNA for the same objects.}\label{fig:s0class}
\end{figure*}

\begin{figure*}
\centering
\includegraphics[width=0.45\textwidth]{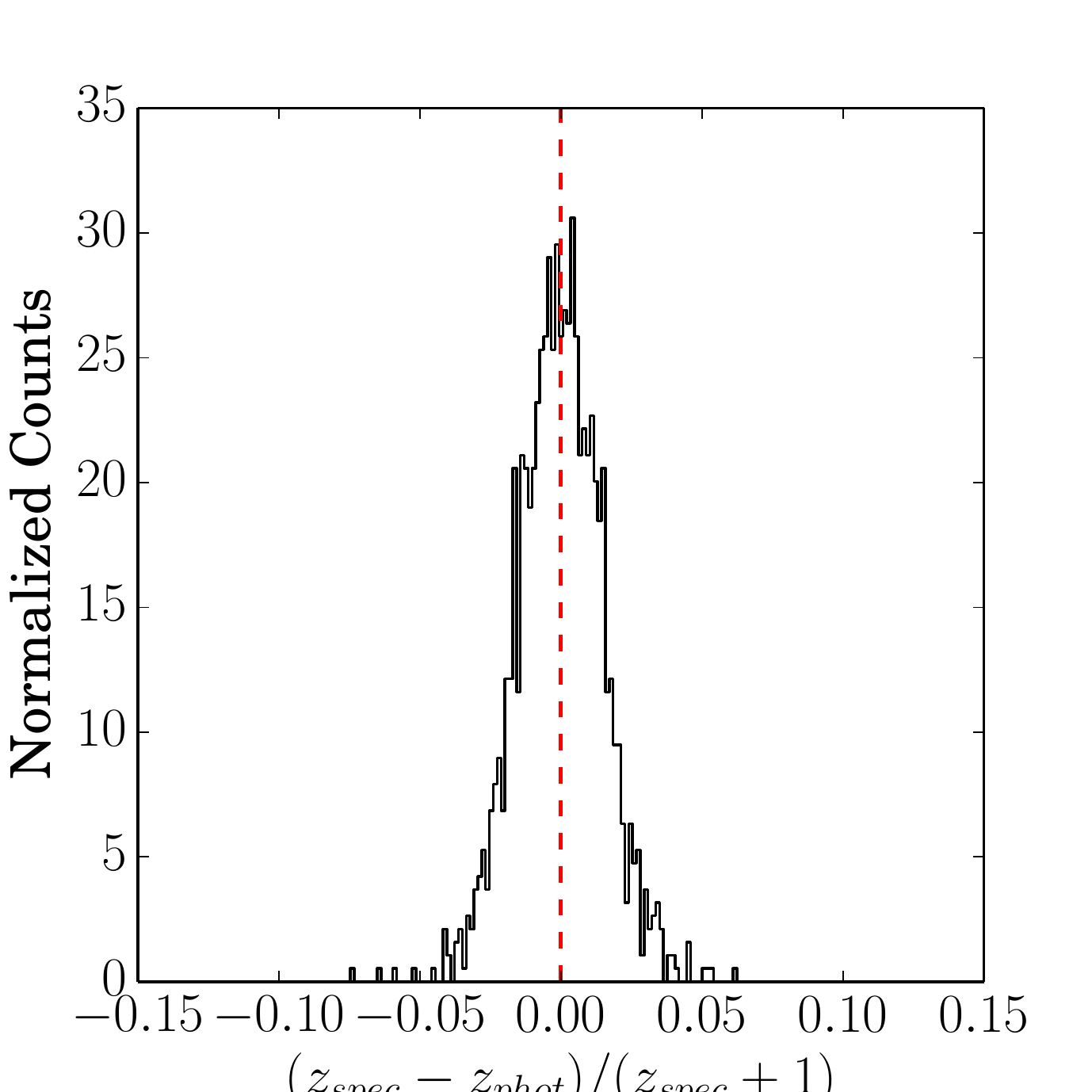} (a)
\includegraphics[width=0.45\textwidth]{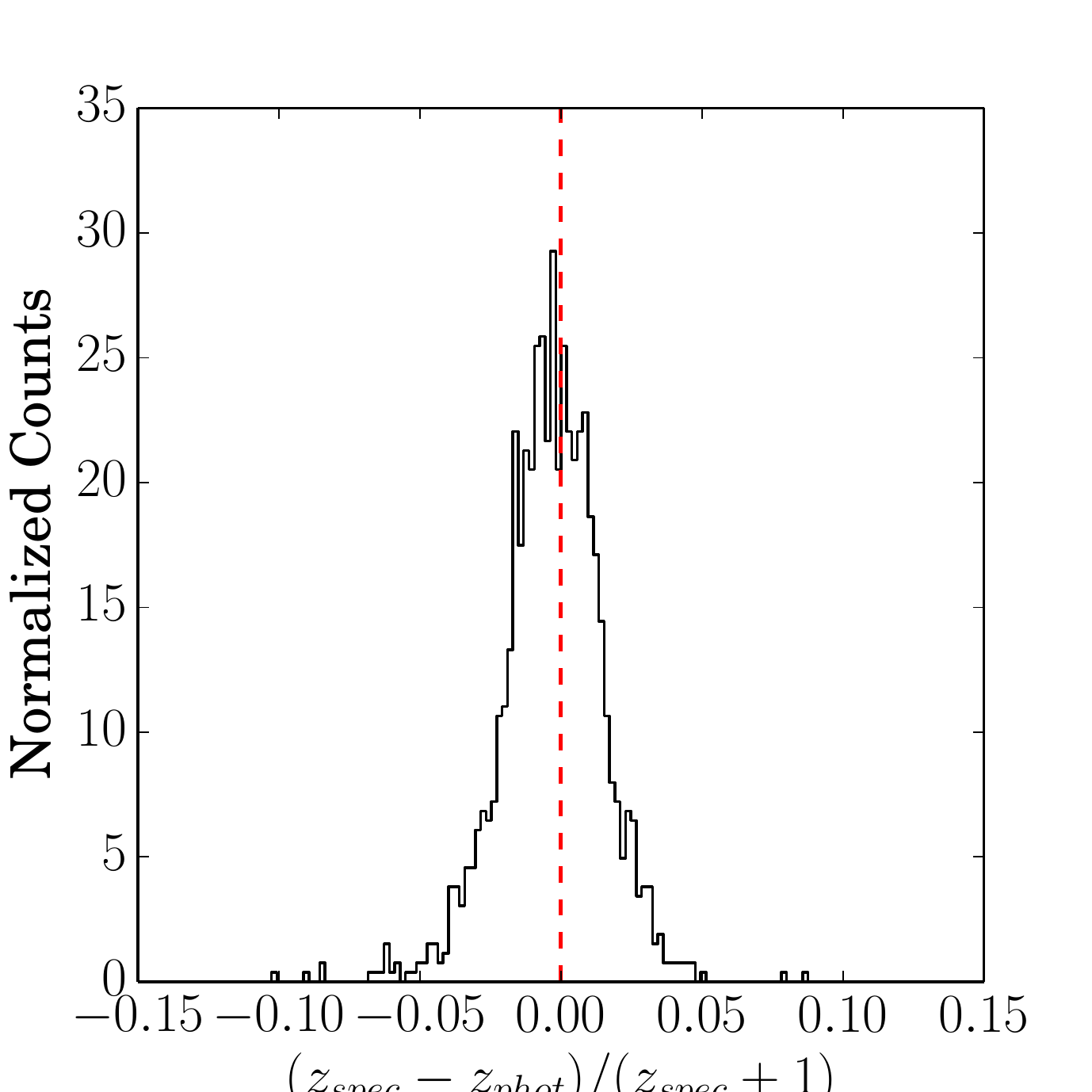} (b)

\caption{Histograms of $\Delta z/(1+z)$ in the case of \textit{Sab} class: left panel represents the results obtained by the \emph{expert} MLPQNA regressor through the proposed workflow, while the right panel represents the result obtained by the standard MLPQNA for the same objects.}
\label{fig:sabclass}
\end{figure*}

\begin{figure*}
\centering
\includegraphics[width=0.45\textwidth]{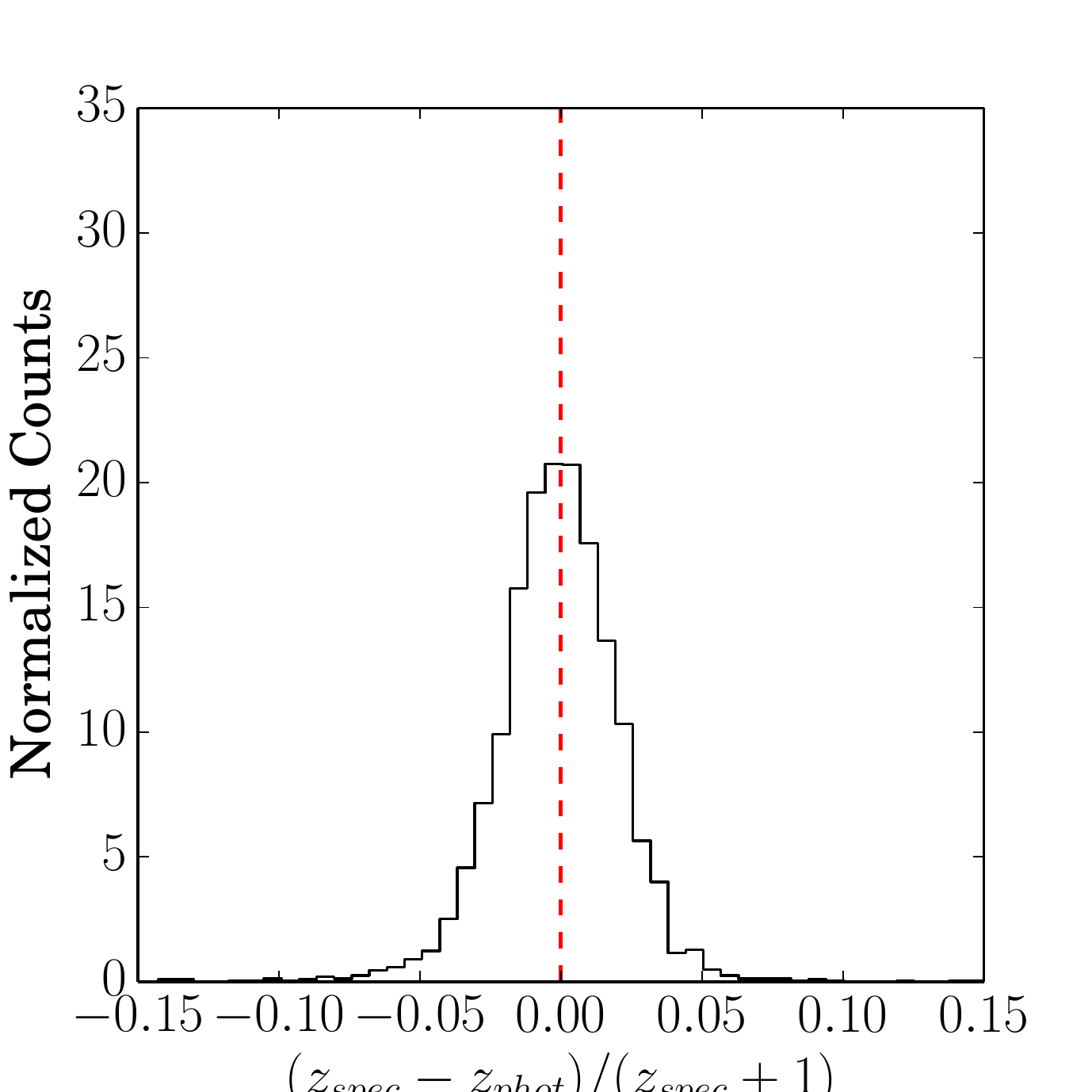} (a)
\includegraphics[width=0.45\textwidth]{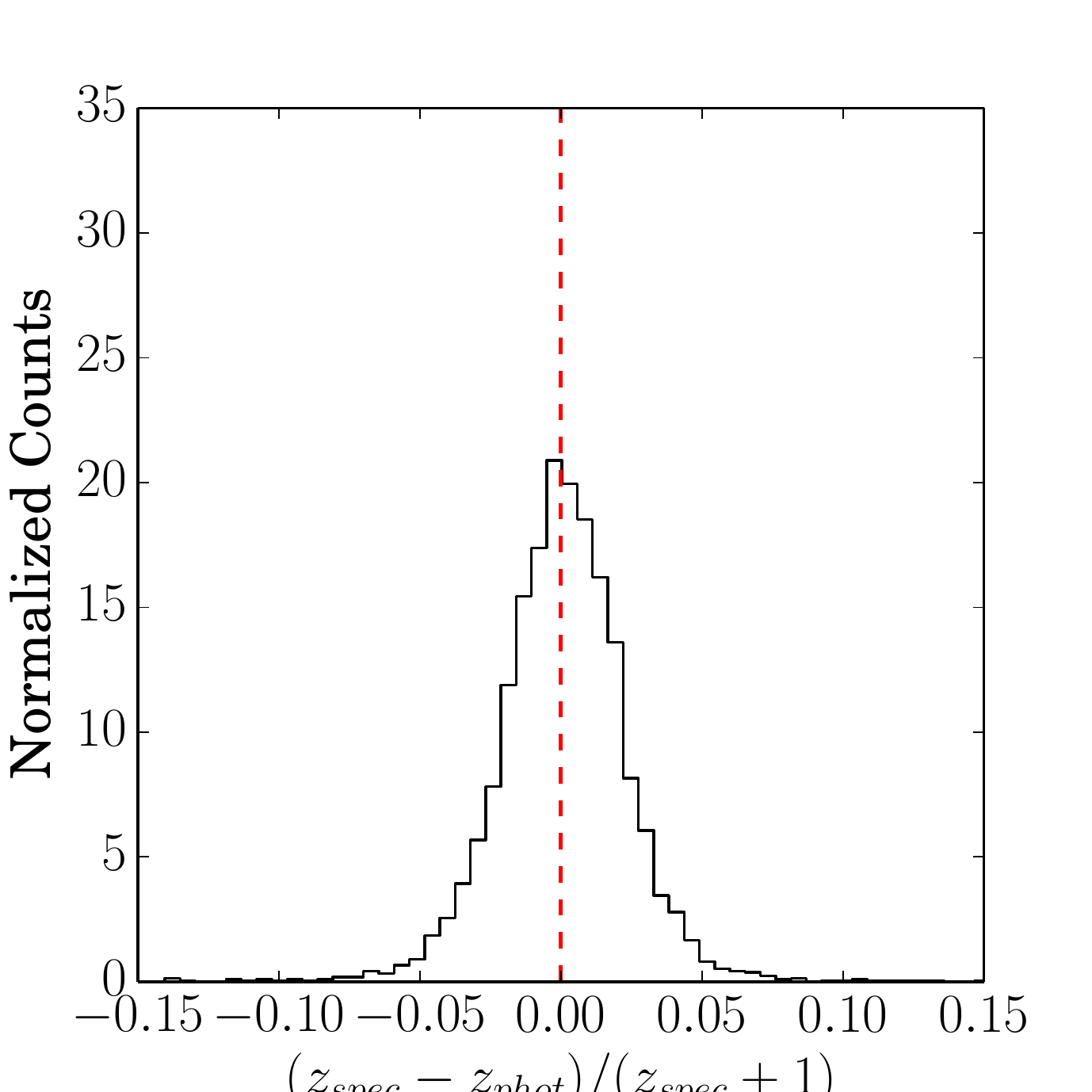} (b)

\caption{Histograms of $\Delta z/(1+z)$ in the case of \textit{Scd} class: left panel represents the results obtained by the \emph{expert} MLPQNA regressor through the proposed workflow, while the right panel represents the result obtained by the standard MLPQNA for the same objects.}\label{fig:scdclass}
\end{figure*}

\begin{figure*}
\centering
\includegraphics[width=0.45\textwidth]{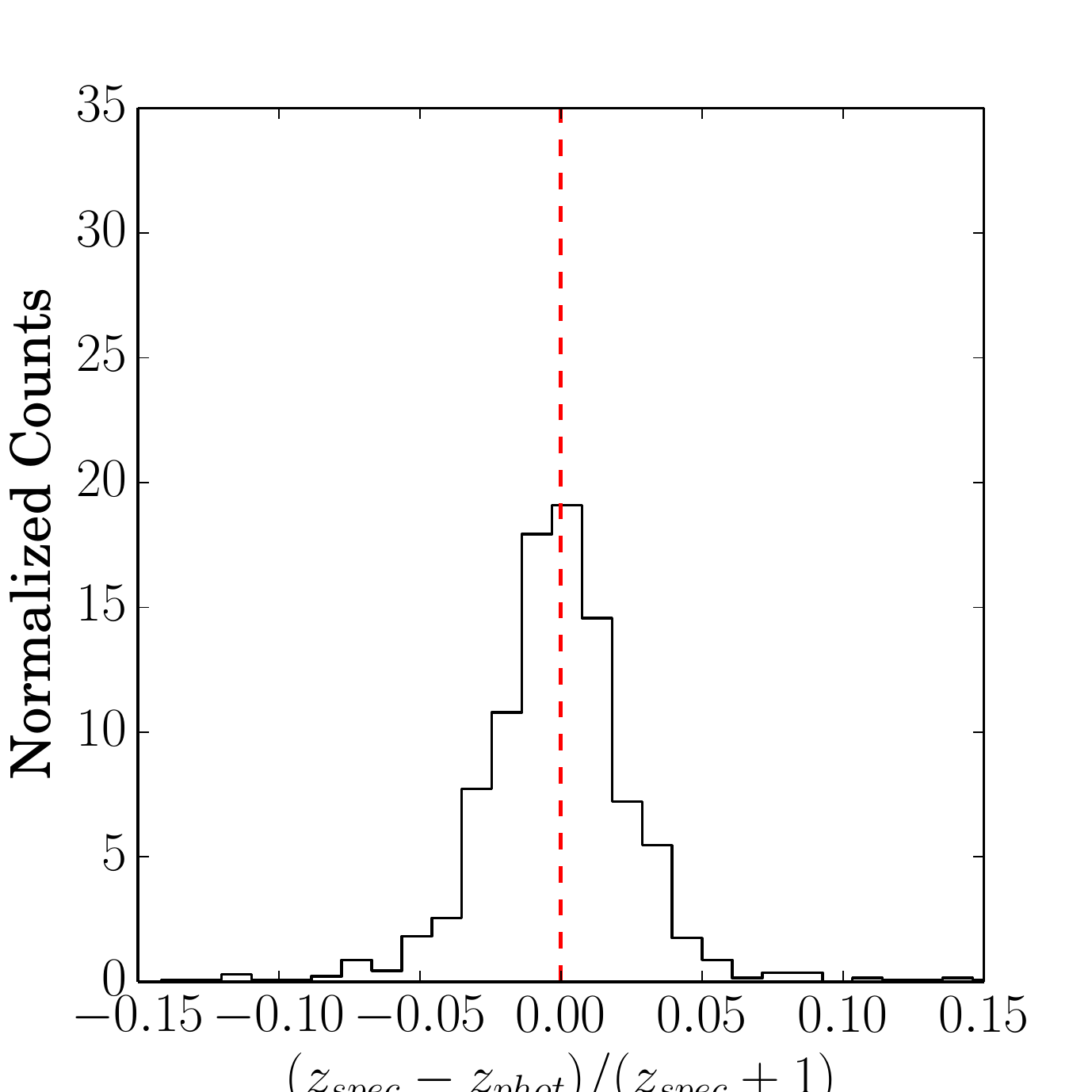} (a)
\includegraphics[width=0.45\textwidth]{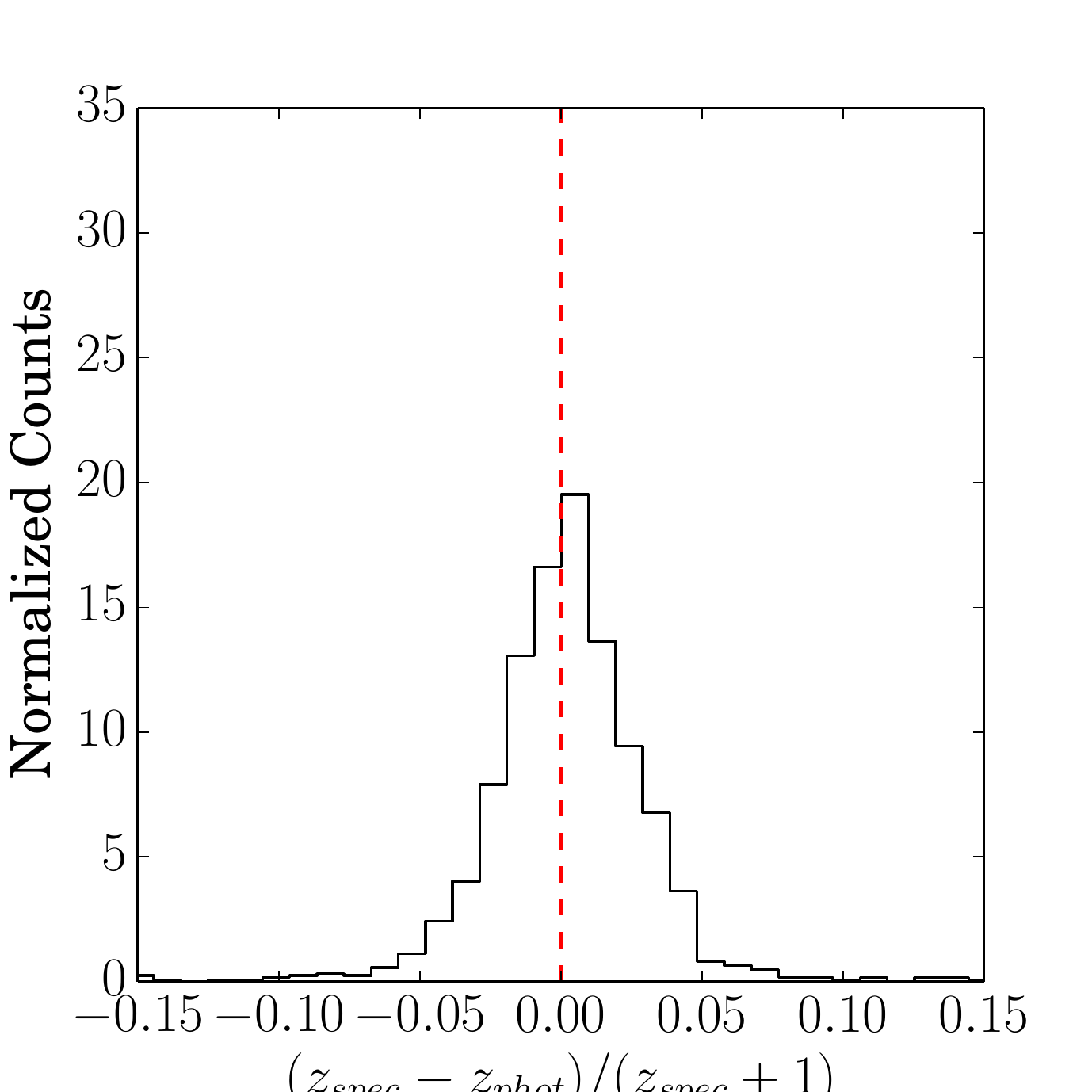} (b)

\caption{Histograms of $\Delta z/(1+z)$ in the case of \textit{SB} class: left panel represents the results obtained by the \emph{expert} MLPQNA regressor through the proposed workflow, while the right panel represents the result obtained by the standard MLPQNA for the same objects.}\label{fig:sbclass}
\end{figure*}

\begin{figure*}
\centering
\includegraphics[width=0.45\textwidth]{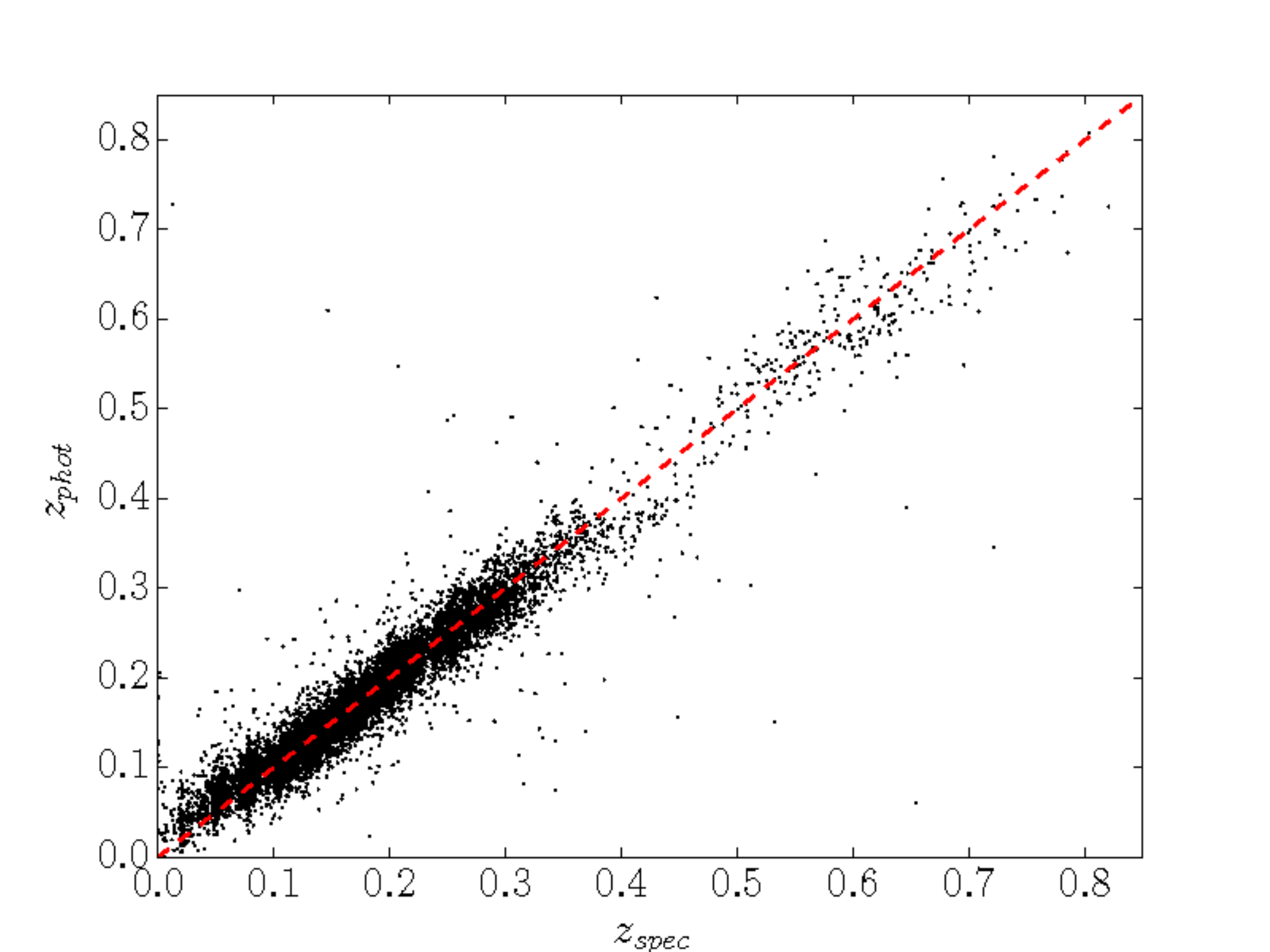} (a)
\includegraphics[width=0.45\textwidth]{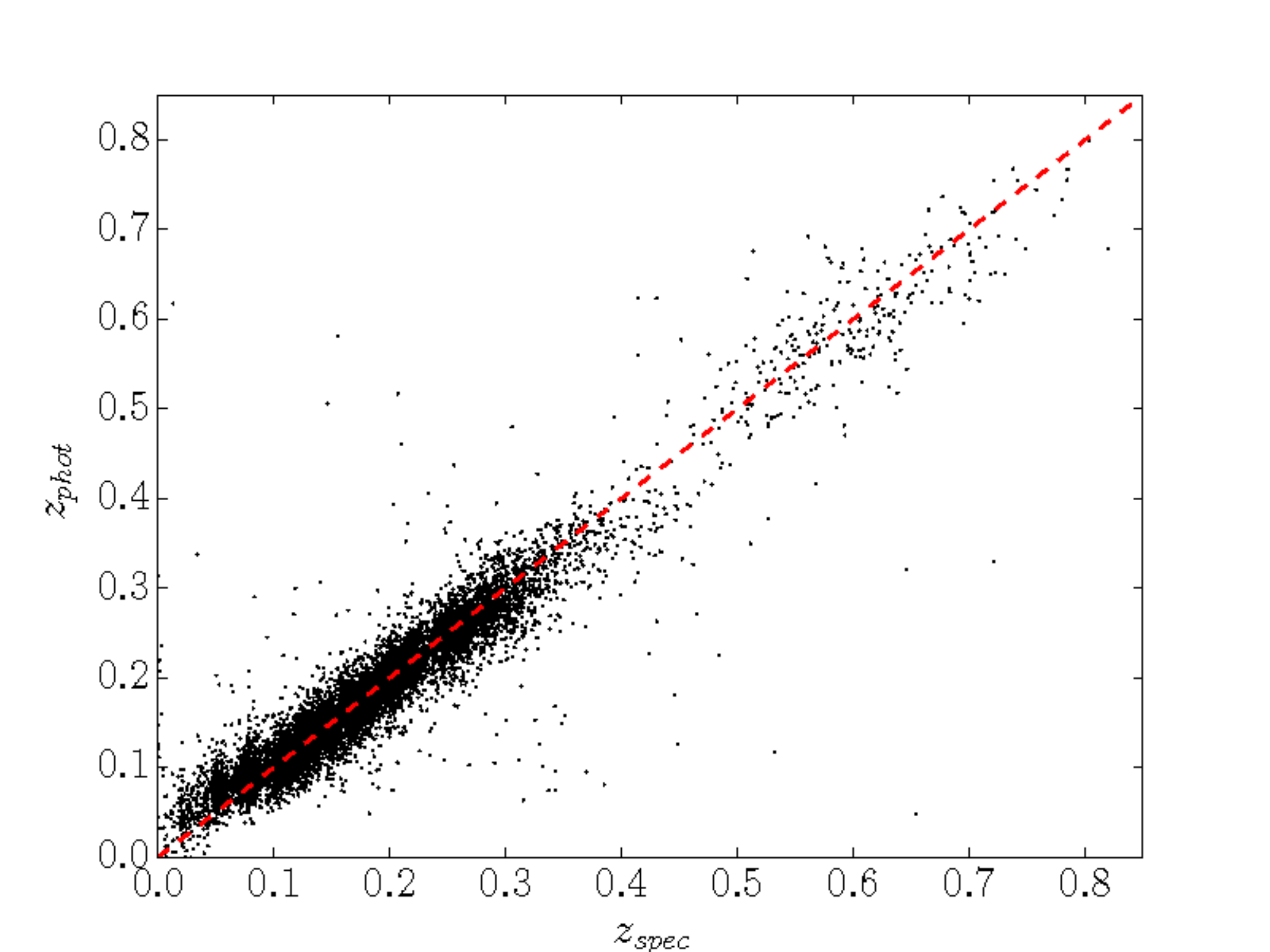} (b)
\includegraphics[width=0.45\textwidth]{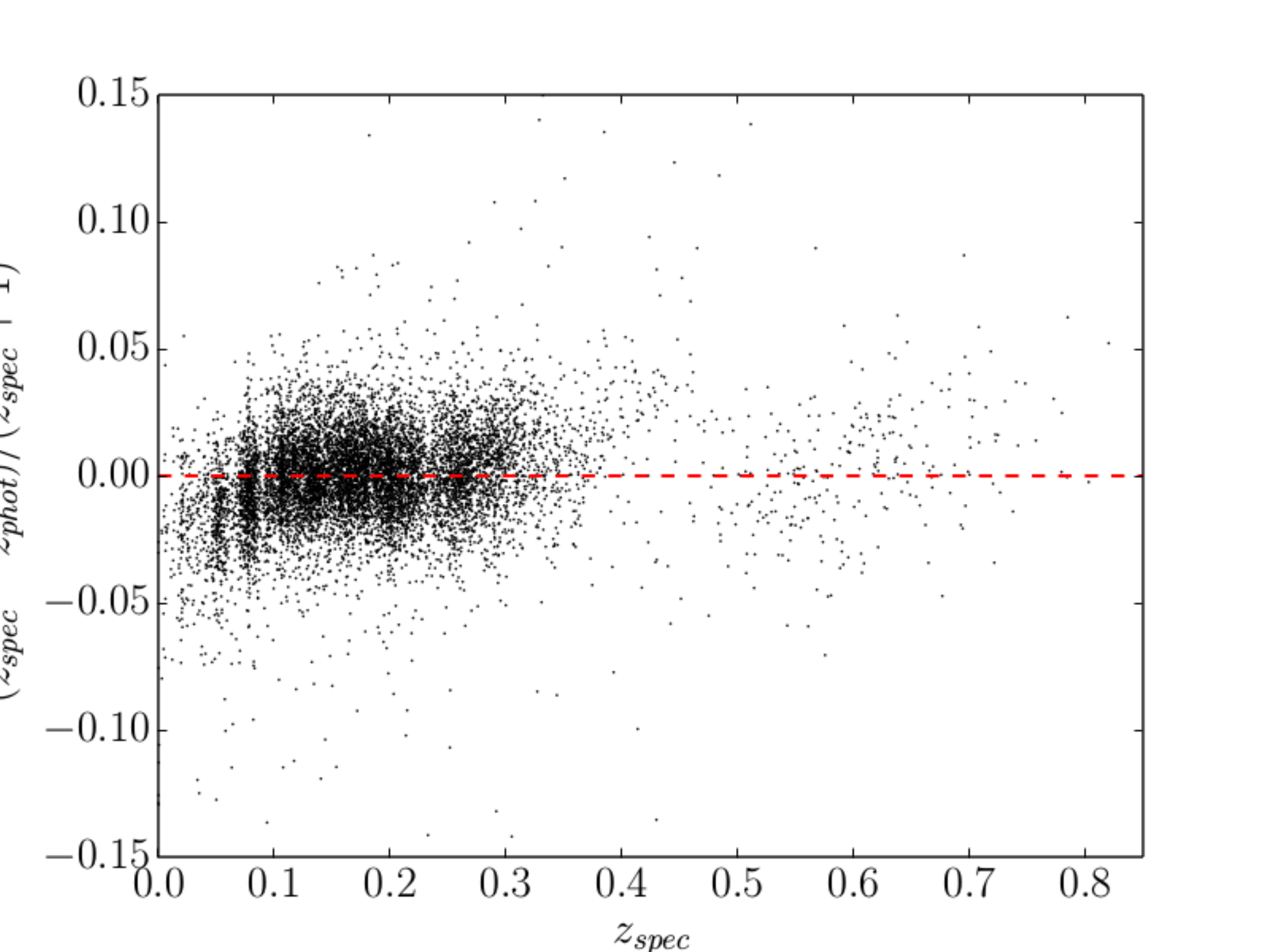} (c)
\includegraphics[width=0.45\textwidth]{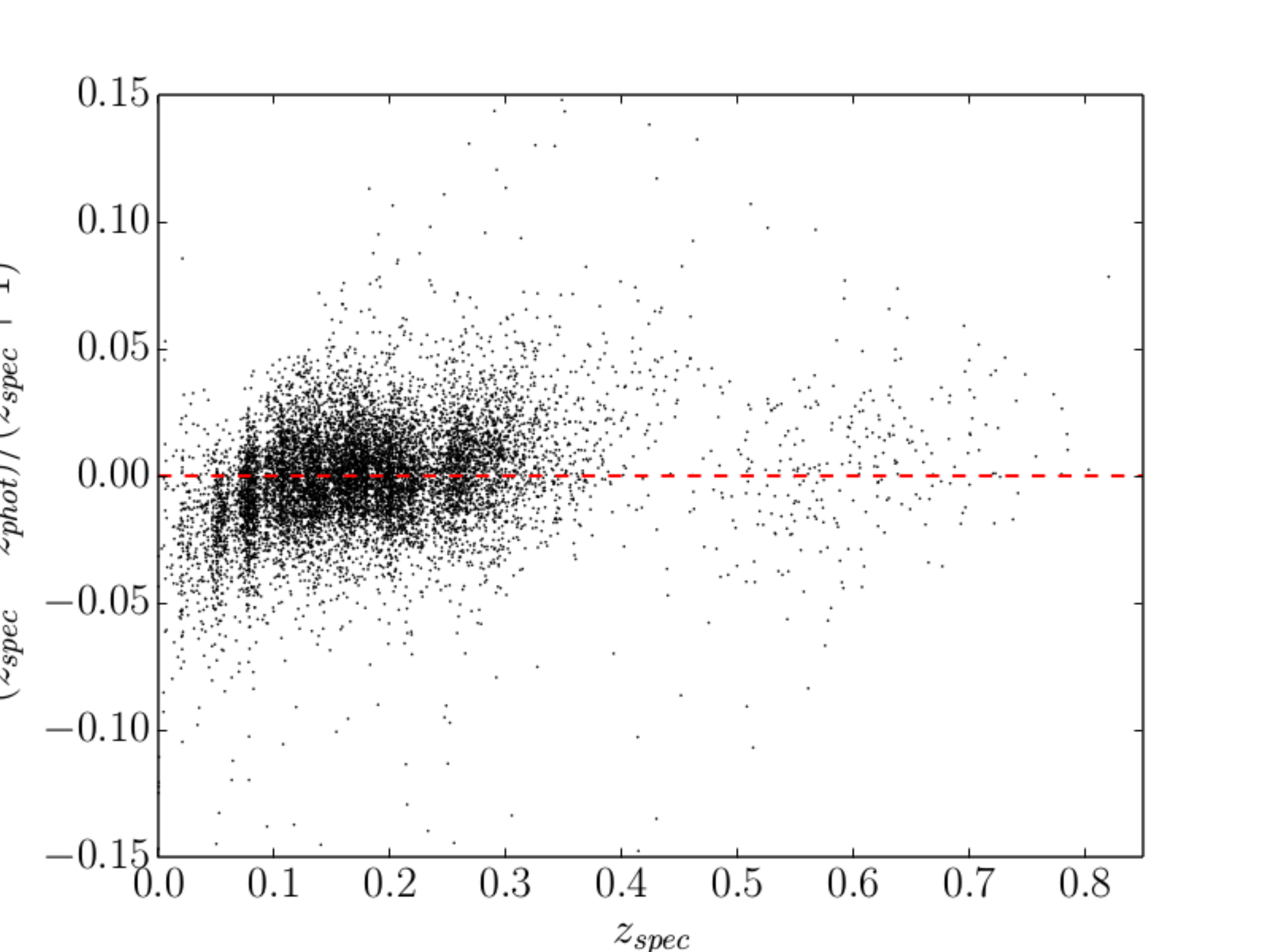} (d)
\includegraphics[width=0.45\textwidth]{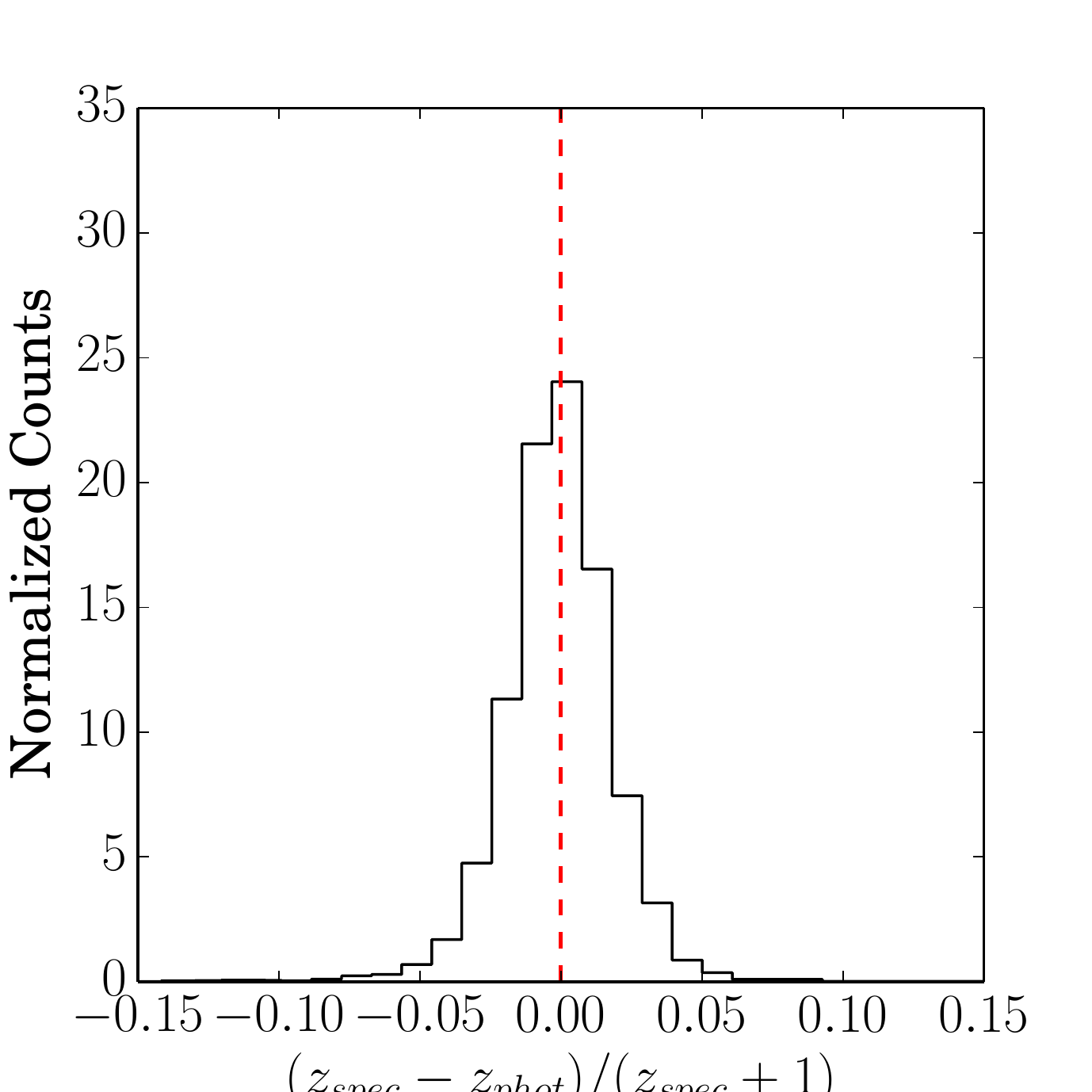} (e)
\includegraphics[width=0.45\textwidth]{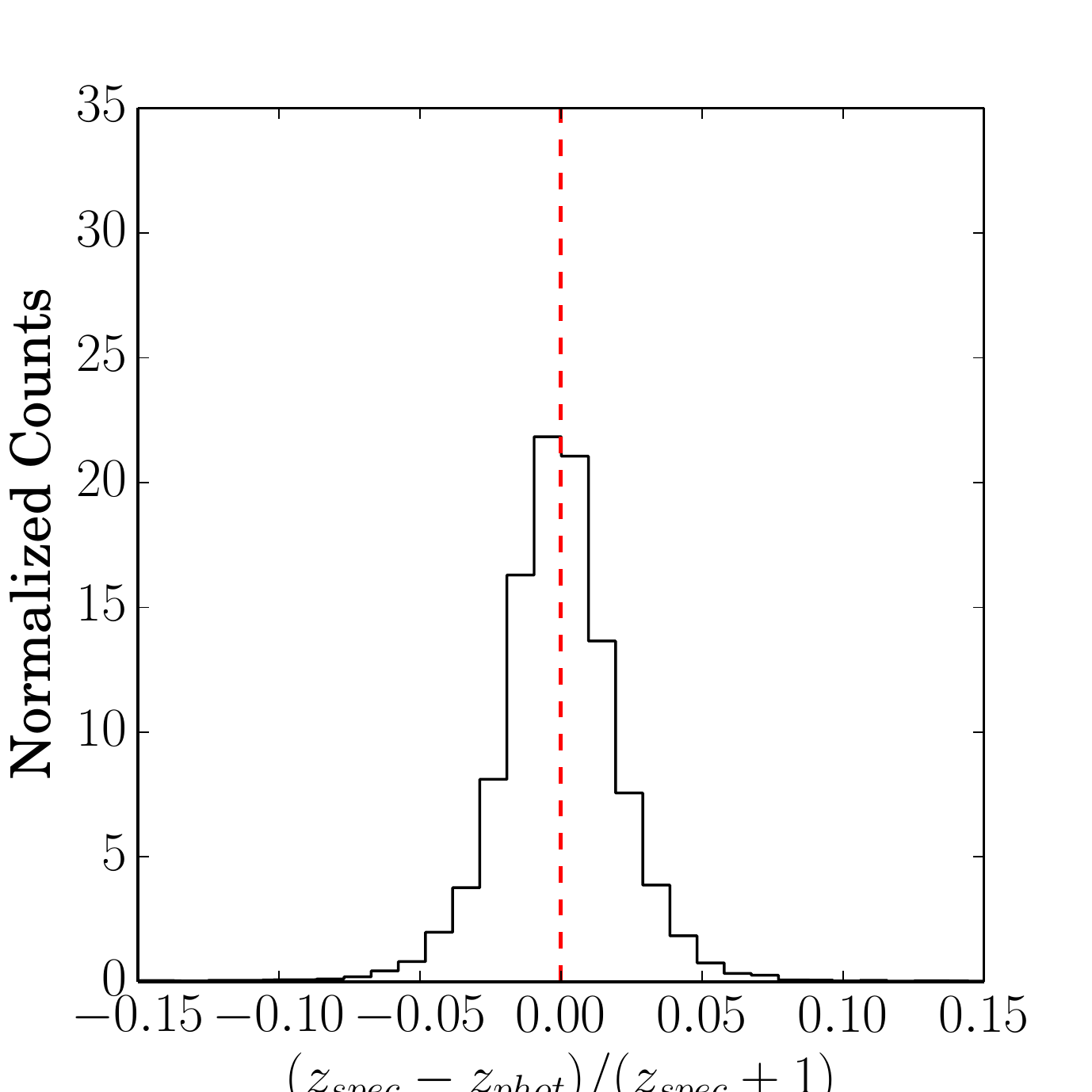} (f)

\caption{The whole set of blind test objects: left column of panels represents the results obtained by grouping together all single spectral-type class outcomes of the \emph{expert} MLPQNA regressors through the proposed \textit{hybrid} workflow, while the right column of panels represents the result obtained by the \textit{standard} MLPQNA for the same objects. The first row shows the diagrams of \zs\ vs. \zp\ ; the second row shows $\Delta z/(1+z)$ vs. $\zs$ diagrams, while the third row shows the histograms of $\Delta z/(1+z)$.}\label{fig:allclass}
\end{figure*}

\label{lastpage}
\end{document}